\documentclass[11pt]{article}
\pdfoutput=1
\usepackage{fullpage}
\usepackage{cite}
\usepackage{graphicx}
\usepackage{amsmath}
\usepackage{amssymb}
\usepackage{subcaption}
\title{}
\date{}

\def\beq{\begin{equation}}
\def\eeq{\end{equation}}
\allowdisplaybreaks
\begin{document}
\bibliographystyle{utphys}
\newcommand{\msbar}{\ensuremath{\overline{\text{MS}}}}
\newcommand{\DIS}{\ensuremath{\text{DIS}}}
\newcommand{\abar}{\ensuremath{\bar{\alpha}_S}}
\newcommand{\bb}{\ensuremath{\bar{\beta}_0}}
\newcommand{\rc}{\ensuremath{r_{\text{cut}}}}
\newcommand{\Nd}{\ensuremath{N_{\text{d.o.f.}}}}
\setlength{\parindent}{0pt}

\titlepage
\begin{flushright}
QMUL-PH-18-20\\
UCLA-18-TEP-107
\end{flushright}

\vspace*{0.5cm}

\begin{center}
{\bf \Large The self-dual classical double copy, and the Eguchi-Hanson
  instanton}

\vspace*{1cm} \textsc{David S. Berman$^a$\footnote{d.s.berman@qmul.ac.uk},
  Erick Chac\'{o}n$^b$\footnote{echacon@fis.cinvestav.mx}, 
Andr\'{e}s Luna$^c$\footnote{andreslunagodoy@hotmail.com} and Chris
  D. White$^a$\footnote{christopher.white@qmul.ac.uk}} \\

\vspace*{0.5cm} $^a$ Centre for Research in String Theory, School of
Physics and Astronomy, \\
Queen Mary University of London, 327 Mile End
Road, London E1 4NS, UK\\

\vspace*{0.5cm} $^b$ Departamento de F\'{i}sica, CINVESTAV-IPN,\\
 Apartado Postal 14740, 07360 M\'{e}xico\\

\vspace*{0.5cm} $^c$ Mani L. Bhaumik Institute for Theoretical
Physics, Department of Physics and Astronomy, University of California
at Los Angeles, California 90095\\

\end{center}

\vspace*{0.5cm}

\begin{abstract}
The double copy is a map from non-abelian gauge theories to gravity,
that has been demonstrated both for scattering amplitudes and exact
classical solutions. In this study, we reconsider the double copy for
exact solutions that are self-dual in either the gauge or gravity
theory. In this case, one may formulate a general double copy in terms
of a certain differential operator, which generates the gauge and
gravity solutions from a harmonic function residing in a biadjoint
scalar theory. As an illustration, we examine the single copy of the
well-known Eguchi-Hanson instanton in gravity. The gauge field thus
obtained represents an abelian-like object whose field is dipole-like
at large distances, and which has no magnetic or electric charge.
\end{abstract}

\vspace*{0.5cm}

\section{Introduction}
\label{sec:intro}

Classical and quantum aspects of various field theories continue to be
intensely studied, due to a plethora of applications in astro-, high
energy and condensed matter physics. Particularly interesting in
recent times have been newly discovered correspondences between
different theories, whose domains of application in nature are widely
separated. In this paper, we focus on the {\it double
  copy}~\cite{Bern:2008qj,Bern:2010ue,Bern:2010yg}, a correspondence
between gauge theories and gravity~\footnote{Throughout, we use the
  term {\it gauge theory} to refer to abelian or non-abelian gauge
  theories, as distinct from gravity, unless otherwise stated.}. In
its original form, the double copy relates perturbative scattering
amplitudes in the two types of theory, subject to a certain mirroring
of colour and kinematic information, {\it BCJ
  duality}~\cite{Bern:2008qj}, being made manifest in the gauge
theory. The relationship is proven to hold at tree
level~\cite{BjerrumBohr:2009rd,Stieberger:2009hq,Bern:2010yg,BjerrumBohr:2010zs,Feng:2010my,Tye:2010dd,Mafra:2011kj,Monteiro:2011pc,BjerrumBohr:2012mg},
where it has a string theoretic
explanation~\cite{Kawai:1985xq}. Although a full loop level proof has
not been found, there is highly non-trivial evidence that the double
copy works at higher orders in perturbation theory in a wide variety
of theories (with and without
supsersymmetry)~\cite{Bern:2010ue,Bern:1998ug,Green:1982sw,Bern:1997nh,Carrasco:2011mn,Carrasco:2012ca,Mafra:2012kh,Boels:2013bi,Bjerrum-Bohr:2013iza,Bern:2013yya,Bern:2013qca,Nohle:2013bfa,
  Bern:2013uka,Naculich:2013xa,Du:2014uua,Mafra:2014gja,Bern:2014sna,
  Mafra:2015mja,He:2015wgf,Bern:2015ooa,
  Mogull:2015adi,Chiodaroli:2015rdg,Bern:2017ucb,Johansson:2015oia},
including all-order evidence in some
cases~\cite{Oxburgh:2012zr,White:2011yy,Melville:2013qca,Luna:2016idw,Saotome:2012vy,Vera:2012ds,Johansson:2013nsa,Johansson:2013aca}. Another
relationship, the {\it zeroth copy}, relates amplitudes in gauge
theory to those in a {\it biadjoint scalar theory}, i.e. containing a
scalar field interacting via two different types of colour charge.\\

The highly non-trivial nature of the above results has prompted many
to ponder whether the double and zeroth copies are an accident of
scattering amplitudes in perturbation theory, or instead indicative of
a deeper -- and hitherto undiscovered -- connection between biadjoint
scalar, gauge and gravity theories. Indeed, the double and zeroth
copies can be extended to a class of exact classical solutions. The
best understood family is that of time-independent Kerr-Schild metrics
in gravity~\cite{Monteiro:2014cda}, where the graviton field has a
highly specific form involving an outer product of a single null
vector. Extensions to this class have been considered
in~\cite{Luna:2015paa,Luna:2016due}, where the single Kerr-Schild and
/ or time independence properties can be at least partially relaxed
(see also ref.~\cite{Goldberger:2016iau} for a discussion of whether
the source terms in the gauge and gravity theory are related, in
addition to the fields). Non-exact classical solutions have also been
examined:
refs.~\cite{Anastasiou:2014qba,Borsten:2015pla,Anastasiou:2016csv,Anastasiou:2017nsz,Cardoso:2016ngt,Borsten:2017jpt,Anastasiou:2017taf,Anastasiou:2018rdx,LopesCardoso:2018xes}
set up a broad programme for identifying linearised fields in a wide
variety of exotic theories, and in a gauge-invariant
manner. Furthermore, classical scattering processes can be double
copied order-by-order in perturbation
theory~\cite{Goldberger:2016iau,Goldberger:2017frp,Goldberger:2017vcg,Goldberger:2017ogt,Luna:2016hge,Luna:2017dtq,Shen:2018ebu,Levi:2018nxp,Plefka:2018dpa,Cheung:2018wkq,Carrillo-Gonzalez:2018pjk},
which promises to vastly streamline the calculation of astrophysically
relevant observables in classical General Relativity, such as those
related to gravitational waves. In a wider context, the double and
zeroth copies have a natural embedding in the so-called CHY
equations~\cite{Cachazo:2013hca,Cachazo:2013iea}, which themselves are
obtainable from ambitwistor string theory~\cite{Mason:2013sva}. These
and other considerations have recently led to extensions of the copies
to curved
spacetimes~\cite{Adamo:2017nia,Bahjat-Abbas:2017htu,Carrillo-Gonzalez:2017iyj},
and to double field theory~\cite{Lee:2018gxc}. Recently, the double
copy has been used to motivate the algebraic classification of
higher-dimensional gravity solutions~\cite{Monteiro:2018xev}.\\

Despite these ongoing efforts, a full understanding of the scope and
generality of the double copy remains elusive. This underlines the
importance of having different ways of thinking about the copy where
it applies, and of looking at different sectors of gauge and gravity
theories for which the copy is particularly clear. One such sector is
that of solutions to the equations of the motion where the field is
{\it self-dual}, which has the effect of projecting out one of the
physical polarisation states of the gluon or graviton. It was shown
some time ago that the amplitude double copy has a particularly
natural form in the self-dual
sector~\cite{Monteiro:2011pc,BjerrumBohr:2012mg,Monteiro:2013rya},
with the added bonus that the BCJ duality property required for gauge
theory amplitudes to satisfy the double copy can be fully
interpreted. This already suggests that a reexamination of the
self-dual sector could be fruitful for gaining new
insights. Furthermore, the first paper to explore the double copy of
exact classical solutions~\cite{Monteiro:2014cda} pointed out that the
properties of self-dual amplitudes can be recast in a Kerr-Schild
language, but where the null vectors are interpreted as differential
operators. The motivation for exploring the self-dual sector in
ref.~\cite{Monteiro:2014cda} was to demonstrate that the Kerr-Schild
copy for classical solutions was merely a different manifestation of
the previously described double copy for amplitudes of
refs.~\cite{Bern:2008qj,Bern:2010ue,Bern:2010yg}. However,
implications of the self-dual Kerr-Schild double copy for classical
solutions were not followed up, and it is the aim of this paper to
explore this in more detail.\\

Remarkably, a related story for exact classical solutions goes all the
way back to ref.~\cite{Tod}, which realised that null electromagnetic
fields could be associated with self-dual Kerr-Schild gravity
metrics. We will review this argument in what follows, interpreting
the results from a modern viewpoint. More specifically, we will see
that given a harmonic scalar function, we may apply a particular
vector differential operator to generate a self-dual solution to the
Maxwell equations. A second application of the same operator generates
a Kerr-Schild graviton, and thus one generates a ladder of fields
residing in biadjoint scalar, gauge and gravity theories. That this is
consistent with the BCJ double copy for amplitudes follows from the
results of ref.~\cite{Monteiro:2014cda}.\\

As an example of the procedure, we examine a particular harmonic
function (first studied in ref.~\cite{Tod}) that gives rise to the
Eguchi-Hanson
solution~\cite{Eguchi:1978xp,Eguchi:1978gw,Eguchi:1979yx}, which
enjoys a well-known interpretation as a gravitational instanton. We
use the self-dual Kerr-Schild approach to study the single copy of
this solution. The single copy is a solution of the (abelian) Maxwell
equations, whose electric and magnetic fields are dipole-like at large
distances. Unlike previously considered solutions, however, there is
no magnetic charge, nor (by self-duality) any electric charge anywhere
in space. Furthermore, the fact that the single copy is abelian-like
rather than non-abelian is itself interesting, given that one might
have expected the single copy of the Eguchi-Hanson instanton to be a
non-abelian instanton e.g. the BPST instanton of
ref.~\cite{Belavin:1975fg}. Parallels between these instanton
solutions were noted even at the inception of the Eguchi-Hanson
solution~\cite{Eguchi:1978xp} (see e.g. ref.~\cite{Ortin:2015hya} for
a more modern review, and refs.~\cite{Lee:2012rb,Lee:2018czs} for
studies in the context of noncommutative theories). Our results will
thus have interesting implications for attempts to study
non-perturbative aspects of the double copy, including those that
attempt to identify symmetries and / or topological invariants in
gauge and gravity theories. \\

The structure of our paper is as follows. We briefly review the
Kerr-Schild double copy in section~\ref{sec:KS}, including aspects
relating to the self-dual sector. In section~\ref{sec:EH}, we examine
in detail the single and zeroth copies of the Eguchi-Hanson solution,
including a discussion of topological properties. We discuss our
results and conclude in section~\ref{sec:conclude}.

\section{Self-dual Kerr-Schild solutions}
\label{sec:KS}

In this section, we briefly review the properties of the Kerr-Schild
double copy, including its formulation tailored to self-dual
solutions. More details can be found in
ref.~\cite{Monteiro:2014cda}. We start by considering a solution
$g_{\mu\nu}$ to Einstein gravity. Expanding around flat space, we may
define a graviton field $h_{\mu\nu}$ via
\begin{equation}
g_{\mu\nu}=\eta_{\mu\nu}+\kappa h_{\mu\nu},
\label{hdef}
\end{equation}
where $\eta_{\mu\nu}$ is the Minkowski metric, and $\kappa^2=16\pi
G_N$, with $G_N$ the Newton constant. The metric $g_{\mu\nu}$ is of
{\it Kerr-Schild form} if the graviton field can be written as
\begin{equation}
h_{\mu\nu}=\Phi\, k_\mu\,k_\nu,
\label{KSdef}
\end{equation}
where $\Phi$ is a scalar field, and $k_\mu$ is null and geodesic:
\begin{equation}
k^2=0,\quad k\cdot\partial k^\mu=0.
\label{kprops}
\end{equation}
Upon substituting the ansatz of eqs.~(\ref{hdef}--\ref{kprops}) into
the Einstein equations, one finds that the Ricci tensor ${R^\mu}_\nu$
truncates at linear order in $\kappa$, so that the graviton obtained
from $\Phi$ and $k_\mu$ constitutes an exact gravitational
solution. Reference~\cite{Monteiro:2014cda} proved that, given any
time-independent Kerr-Schild graviton, the vector field
\begin{equation}
A^a_\mu=c^a\Phi\,k_\mu
\label{AmuKS}
\end{equation}
(where $c^a$ is an arbitrary colour vector) is a solution of the
abelian Yang-Mills (Maxwell) equations
\begin{equation}
\partial^\mu F^a_{\mu\nu}=0.
\label{Maxwell}
\end{equation}
Thus, any such gravity solution has a well-defined gauge theory
counterpart, which can be interpreted as the ``single copy'' of the
former. Furthermore, upon repeating the procedure of replacing one of
the null vectors $k^\mu$ with an arbitrary colour vector, one obtains
the single scalar field 
\begin{equation}
\Phi^{aa'}=c^a\tilde{c}^{a'}\Phi,
\label{Phiaa'}
\end{equation} 
that is found to satisfy the linearised biadjoint scalar field
equation
\begin{equation}
\partial^2\Phi^{aa'}=0.
\label{biadjoint}
\end{equation}
Thus, the field $\Phi^{aa'}$ is to be regarded as the ``zeroth copy''
associated with the gravity solution $h_{\mu\nu}$ and gauge field
$A_\mu$, and a number of non-trivial examples of this correspondence
were given in ref.~\cite{Monteiro:2014cda}, including the
Schwarzschild and Kerr black holes. Extensions to include multiple
Kerr-Schild-like terms were considered in ref.~\cite{Luna:2015paa},
for which the Taub-NUT solution~\cite{Taub,NUT} was found to be a
special case. Time dependence was also added in
ref.~\cite{Luna:2016due}, which considered an arbitrarily accelerating
point charge (mass) in gauge theory (gravity). Known amplitudes for
photon and graviton Bremstrahlung were recovered as part of this
picture, making clear that the Kerr-Schild double copy for classical
solutions is a manifestation of the same double copy that relates
scattering amplitudes in biadjoint scalar, gauge and gravity
theories.\\

In this paper, we focus on solutions that are {\it self-dual}, namely
those satisfying~\footnote{Equations~(\ref{selfdualgauge})
  and~(\ref{selfdualgrav}) are presented for Lorentzian signature in four dimensions. In the Euclidean case that follows, these equations must be modified with the $i$ being absent.}
\begin{equation}
F_{\mu\nu}=\tilde{F}_{\mu\nu}\equiv
\frac{i}{2}  \sqrt{{\rm{det}}|g|}  \epsilon_{\mu\nu\rho\sigma}F^{\rho\sigma}
\label{selfdualgauge}
\end{equation}
and
\begin{equation}
R_{\mu\nu\lambda\sigma}=\tilde{R}_{\mu\nu\lambda\sigma}\equiv
\frac{i}{2} \sqrt{{\rm{det}}|g|} \epsilon_{\mu\nu\rho\sigma}
{R^{\rho\sigma}}_{\lambda\sigma}
\label{selfdualgrav}
\end{equation}
in gauge theory and gravity respectively, where $F_{\mu\nu}$
($R_{\mu\nu\lambda\sigma}$) is the field strength tensor (Riemann
curvature), and $\epsilon_{\mu\nu\rho\sigma}$ the Levi-Cevita
symbol. Physically, such solutions correspond to keeping only one of
the two physical polarisation states of the photon or graviton, and
for such solutions one may think about the Kerr-Schild double copy in
a different way. Namely, one may formulate the Kerr-Schild vector
$k_\mu$ as a differential operator $\hat{k}_\mu$, which amounts to
considering it as multiplicative in momentum rather than position
space. The symmetry of the resulting graviton
\begin{equation}
h_{\mu\nu}=\hat{k}_\mu\,\hat{k}_\nu\,\Phi
\label{gravKS2}
\end{equation}
then implies that $[\hat{k}_\mu,\hat{k}_\nu]=0$ i.e. $\hat{k}_\mu$
commutes with itself. Furthermore, we may implement the null and
geodesic Kerr-Schild conditions as
\begin{equation}
\eta_{\mu\nu}\hat{k}^\mu\hat{k}^\nu=0,\quad \hat{k}\cdot\partial=0 \, .
\label{KSconditions}
\end{equation}
A suitable operator can be defined by adopting the lightcone
coordinates
\begin{equation}
u=\frac{t-iz}{\sqrt{2}},\quad v=\frac{t+iz}{\sqrt{2}},\quad
X=\frac{ix-y}{\sqrt{2}},\quad Y=\frac{ix+y}{\sqrt{2}},
\label{lightcone}
\end{equation}
where, following ref.~\cite{Tod}, we define $(u,v,X,Y)$ in terms of
Euclidean Cartesian coordinates $(t,x,y,z)$ in ${\mathbb R}^4$. The
operator $\hat{k}_\mu$ may now be chosen as
\begin{equation}
\hat{k}_u=\partial_X,\quad k_v=0,\quad k_X=0,\quad k_Y=\partial_v,
\label{khatdef}
\end{equation}
and we may then show that the Einstein equations for the graviton of
eq.~(\ref{gravKS2}) reduce to
\begin{equation}
\partial^2\Phi+\kappa\{\partial_X\Phi,\partial_u\Phi\}=0,
\label{Plebanski}
\end{equation}
where we introduced the Poisson bracket of two functions $f$ and $g$:
\begin{equation}
\{f,g\}=(\partial_X f)(\partial_u g)-(\partial_u f)(\partial_X g) \, .
\label{Poisson}
\end{equation}
Equation~(\ref{Plebanski}) is the {\it Plebanski equation} describing
self-dual gravity, and a similar equation can be obtained for
Yang-Mills theory by considering the Kerr-Schild single copy
field
\begin{equation}
A_\mu^a=\hat{k}_\mu \Phi^a,
\label{Amuadef}
\end{equation}
for an adjoint-valued scalar field $\Phi=\Phi^a {\bf T}^a$, where
${\bf T}^a$ is a generator of the gauge group. With this ansatz, the
Yang-Mills equations turn out to imply
\begin{equation}
\partial^2 \Phi+2ig[\partial_v\Phi,\partial_X\Phi]=0,
\label{selfdualYM}
\end{equation}
a known equation\footnote{The form of eq.~(\ref{selfdualYM}) differs
  slightly from that given in
  refs.~\cite{Parkes:1992rz,Monteiro:2013rya}, due to different
  conventions for the lightcone coordinates and Kerr-Schild operator.}
for self-dual Yang-Mills theory first derived in
ref.~\cite{Parkes:1992rz}, and studied within a double copy context in
ref.~\cite{Monteiro:2013rya}, where it was used to argue that
amplitudes in the self-dual sector are manifest double copies of each
other. Furthermore, a kinematic algebra underlying the numerators of
self-dual amplitudes was obtained, that provided an explicit
realisation of the BCJ duality between colour and kinematic degrees of
freedom. The identification of eq.~(\ref{Amuadef}) makes clear that
the Kerr-Schild single copy for self-dual solutions is indeed the same
as the double copy as originally formulated for amplitudes. However,
it can also be noted that eq.~(\ref{Amuadef}) is equally valid for
exact classical solutions. In this context, this approach is actually
a modern reinterpretation of the much earlier ref.~\cite{Tod}, which
also recognised that self-dual Kerr-Schild gravitons could be
associated with null Maxwell fields.\\

We may adapt the argument of ref.~\cite{Tod} to the above discussion
as follows. First, we rewrite the Plebanski equation of
eq.~(\ref{Plebanski}) as
\begin{equation}
\Phi_{,uv}-\Phi_{,XY}=-\Phi_{,vv}\Phi_{,XX}+(\Phi_{,vX})^2,
\label{Plebanski2}
\end{equation}
where the comma denotes the partial derivative. The metric of
eq.~(\ref{gravKS2}) can be written
\begin{equation}
ds^2=2du\,dv-2dX\,dY-2\kappa\left(\Phi_{,XX}du^2+2\Phi_{,vX}du\,dY
+\Phi_{,vv}dY^2\right),
\label{gravKS3}
\end{equation}
where we have used the form of the Minkowski metric
\begin{equation}
\eta_{\mu\nu}=2du\,dv-2dX\,dY
\label{etauv}
\end{equation}
in the lightcone coordinates of eq.~(\ref{lightcone}). However, if the
metric of eq.~(\ref{etauv}) is to be in Kerr-Schild form, then the
${\cal O}(\kappa)$ term must be a perfect square, which in turn
implies
\begin{equation}
\Phi_{,vv}\Phi_{,XX}-(\Phi_{,vX})^2=0.
\label{perfectsquare}
\end{equation}
Coupled with the requirement of self-duality, eq.~(\ref{Plebanski2}),
this implies that the function $\Phi$ must be harmonic. But this is
none other than the requirement that the field $\Phi$ satisfy the
linearised biadjoint scalar equation of eq.~(\ref{biadjoint}). Thus,
we can summarise the results of ref.~\cite{Tod} from a double copy
point of view as follows: given a harmonic function $\Phi$, we may
consider the fields
\begin{displaymath}
\Phi,\quad A_\mu=\hat{k}_\mu\Phi,\quad h_{\mu\nu}=\hat{k}_\mu\,
\hat{k}_\nu\Phi,
\end{displaymath}
living in a biadjoint scalar, gauge and gravity theory
respectively~\footnote{Strictly speaking, we must dress the biadjoint
  and gauge fields with arbitrary colour vectors as in
  eqs.~(\ref{AmuKS}, \ref{Phiaa'}). However, given that the equations
  of motion linearise in each theory for the solutions considered, we
  can ignore this unless otherwise stated.}. The fields $A_\mu$ and
$\Phi$ constitute the classical single and zeroth copies of the
graviton $h_{\mu\nu}$ respectively. Having set up the general
framework for self-dual classical solutions, we now consider a
particular example in detail.

\section{The Eguchi-Hanson solution}
\label{sec:EH}

In the previous section, we described a procedure for generating gauge
theory and graviton fields obeying the double copy, by acting upon a
harmonic function in lightcone coordinates with the differential
operator of eq.~(\ref{khatdef}). In this section, we study a
particular example in detail, namely the {\it Eguchi-Hanson solution}
of Euclidean gravity, originally derived in
refs.~\cite{Eguchi:1978xp}, and interpreted in more detail in
refs.~\cite{Eguchi:1978gw,Eguchi:1979yx,tHooft:1988wxy}. A
conventional way to present this solution is to choose a radial
coordinate $r$ in ${\mathbb R}^4$, and three Euler angles
\begin{equation}
0\leq\theta\leq\pi,\quad 0\leq \phi\leq
2\pi, \quad 0\leq \psi\leq 4\pi,
\label{Euler}
\end{equation}
such that the line element may be written
\begin{equation}
ds^2_{\rm EH}=\left[1-(a/r)^4\right]^{-1}
dr^2+r^2\left\{\sigma_x^2+\sigma_y^2+\left[1-(a/r)^4\right]
\sigma_z^2\right\},
\label{ds2EH}
\end{equation}
where 
\begin{align}
\sigma_x&=\frac12(\sin\psi\,d\theta-\sin\theta\,\cos\psi\,d\phi),\notag\\
\sigma_y&=\frac12(-\cos\psi\,d\theta-\sin\theta\,\sin\psi\,d\phi),\notag\\
\sigma_z&=\frac12(d\psi+\cos\theta\,d\phi).
\label{sigmaxyz}
\end{align}
With this choice of coordinates, the topology of the spacetime is that
of ${\mathbb R}_+\times S^3$, although it has a singularity at
$r=a$. We can characterise the topology in four spacetime dimensions
using two invariants, namely the Hirzebruch signature $\tau$ and the
Euler characteristic $\chi$ (see e.g.~\cite{Eguchi:1980jx} for a
review). These receive contributions from the bulk of the spacetime,
and the boundary, and are both zero for ${\mathbb R}_+\times
S^3$~\cite{Eguchi:1978xp} (n.b. to get this result, it is crucial that
one includes the boundary contribution from the singularity at
$r=a$). \\

However, one can also construct a topologically non-trivial spacetime
from the above metric, by removing the singularity at
$r=a$~\cite{Eguchi:1978gw}. Changing variables to 
\begin{equation}
u=r^2\left[1-(a/r)^4\right],
\label{udef}
\end{equation}
and expanding near $u=0$ ($r=a$), one finds
\begin{equation}
ds_{\rm EH}^2\simeq \frac14 du^2+\frac14 u^2(d\psi+\cos\theta\,d\phi)^2
+\frac{a^2}{4}(d\theta^2+\sin^2\theta\,d\phi^2),
\label{ds2EH2}
\end{equation}
such that the local topology near $r=a$ is that of ${\mathbb
  R}^2\times S^2$, where the latter factor is associated with the
coordinates $(\theta,\pi)$. At fixed values of these coordinates one
has the pure ${\mathbb R}^2$ metric
\begin{equation}
ds_{\rm EH}^2\Big|_{\theta=\phi={\rm const.}}\simeq \frac14
(du^2+u^2d\psi^2),
\label{R2metric}
\end{equation}
whose singularity at $u=0$ is removable provided one restricts the
Euler angle $\psi$ to lie in the range
\begin{equation}
0\leq \psi \leq 2\pi,
\label{psirange}
\end{equation}
rather than the full range of eq.~(\ref{Euler}). Put another way, this
has the effect of identifying each spacetime point $x^\mu$ with
$-x^\mu$, so that the boundary of the spacetime becomes the projective
space $P_3({\mathbb R})=S^3/Z_2$ rather than $S^3$. Recalculation of
the topological invariants mentioned above then yields $\chi=2$,
$\tau=1$. Removable singularities of this type were christened {\it
  bolts} in ref.~\cite{Gibbons:1979xm}, as distinct from {\it
  (anti)-nut} singularities (such as those in the Taub-NUT metric
studied in a double copy context in ref.~\cite{Luna:2015paa}). The
invariants $\chi$ and $\tau$ then have an interpretation in terms of a
superposition of the number of nuts and
bolts~\cite{Gibbons:1979xm}. The spacetime obtained by removing the
bolt singularity in the present case has zero action, is everywhere
regular, and is also self-dual. It has thus been widely studied as an
example of a gravitational instanton. \\

Let us now turn, as promised, to the single copy of the Eguchi-Hanson
solution, for which we first need to know the zeroth copy field
$\Phi$. Following section~\ref{sec:KS}, this is a harmonic function
residing in a biadjoint scalar field theory, that one must act upon
with the Kerr-Schild operator of eq.~(\ref{khatdef}) in order to
produce a gauge field and, subsequently, the Eguchi-Hanson
graviton. Furthermore, this function must be given in terms of the
lightcone coordinates of eq.~(\ref{lightcone}). The relevant function
has been given in ref.~\cite{Tod} as
\begin{equation}
\Phi=\frac{\lambda X^2}{2u^2(uv-XY)},
\label{PhiEH}
\end{equation} 
from which one obtains the gauge field
\begin{equation}
A_\mu=\hat{k}_\mu\Phi=\lambda \frac{X^2}{2u^2(uv-XY)^2}
\left(\frac{2uv-XY}{X},0,0,-u\right)
\label{AmuEH}
\end{equation}
and graviton 
\begin{equation}
h_{\mu\nu}=\lambda \frac{(vdu-XdY)^2}{(uv-XY)^3}.
\label{hmunuEH}
\end{equation}
Here we have included a common parameter $\lambda$, so that vacuum
solutions in each theory correspond to $\lambda\rightarrow 0$. We do
not include explicit factors of the coupling in the gauge or biadjoint
theories, choosing instead to absorb these in the arbitrary colour
vectors that would dress the above solutions. \\

Let us first check that eq.~(\ref{hmunuEH}) indeed corresponds to the
Eguchi-Hanson metric of eq.~(\ref{ds2EH}), which may write more
clearly by substituting eq.~(\ref{sigmaxyz}) into eq.~(\ref{ds2EH}):
\begin{equation}
ds_{\rm EH}^2=\frac14 r^2\left[1-(a/r)^4\right]\left[d\psi+\cos\theta\,d\phi
\right]^2+\left[1-(a/r)^4\right]^{-1}dr^2+\frac14 r^2\left[d\theta^2
+\sin^2\theta\,d\phi^2\right].
\label{EHtrans1}
\end{equation}
This equivalence has been shown using twistor methods in
ref.~\cite{Burnett-Stuart}, but we can instead use a more
straightforward approach. Adding the background flat line element (in
lightcone coordinates) to eq.~(\ref{hmunuEH}), one obtains
\begin{equation}
ds^2=2du\,dv-2dX\,dY+\lambda \frac{(vdu-XdY)^2}{(uv-XY)^3}.
\label{EHtrans2}
\end{equation}
The lightcone coordinates may be transformed to (Euclidean) Cartesian
coordinates according to eq.~(\ref{lightcone}), after which one may
make the further transformation
\begin{align}
x&=r\cos\left(\frac{\theta}{2}\right)\cos\left(\frac{\phi+\psi}
{2}\right),\notag\\
y&=r\cos\left(\frac{\theta}{2}\right)\sin\left(\frac{\phi+\psi}
{2}\right),\notag\\
z&=r\sin\left(\frac{\theta}{2}\right)\cos\left(\frac{\phi-\psi}
{2}\right),\notag\\
t&=r\sin\left(\frac{\theta}{2}\right)\sin\left(\frac{\phi-\psi}
{2}\right),
\label{Polardef}
\end{align}
where the angles have ranges as in eq.~(\ref{Euler}). Here $r$ is the
radial distance in ${\mathbb R}^4$, and from the two sets of
transformations in eqs.~(\ref{lightcone}, \ref{Polardef}) we can
establish the relation
\begin{equation}
r^2=x^2+y^2+z^2+t^2=2(uv-XY).
\label{rdef}
\end{equation}
Next, we can connect the parameter $\lambda$ from eq.~(\ref{hmunuEH})
with the parameter $a$ appearing in the metric of eq.~(\ref{ds2EH}),
by considering the Kretschmann scalar in both the lightcone and polar
coordinate systems:
\begin{equation}
R^{\alpha\beta\mu\nu}R_{\alpha\beta\mu\nu}=\frac{384 a^8}{r^{12}}
=\frac{24\lambda^2}{(uv-XY)^6}. 
\label{Kretschmann}
\end{equation}
Equation~(\ref{rdef}) then immediately implies
\begin{equation}
\lambda=\frac{a^4}{2},
\label{alambda}
\end{equation}
such that transforming eq.~(\ref{EHtrans2}) into the polar coordinate
system of eq.~(\ref{Polardef}) yields 
\begin{align}
ds^2&=dr^2+\frac14 r^2\left[d\psi^2+d\phi^2+d\theta^2+2\cos\theta
d\psi\,d\phi\right]+\frac{4a^4}{r^6}\left(\frac{rdr}{2}-\frac{ir^2}{4}(d\psi
+\cos\theta\,d\phi)\right)^2.
\label{EHtrans3}
\end{align}
Finally, one may perform the additional transformation
\begin{equation}
\psi\rightarrow \psi+\frac{i}{2}\log\left(\frac{r^4-a^4}{r^4}\right)
\label{psitrans}
\end{equation}
to put eq.~(\ref{EHtrans3}) into the form of eq.~(\ref{ds2EH}). Note
that this last transformation is singular, and creates the explicit
coordinate singularity at $r=a$ (the ``bolt'') appearing in
eq.~(\ref{ds2EH}). \\

Armed with the above transformations, we can also transform the zeroth
copy field $\Phi$ and single copy gauge field $A_\mu$ of
eqs.~(\ref{PhiEH}, \ref{AmuEH}) into the polar coordinate system, and
we find
\begin{equation}
\Phi=\frac{r^2}{r^4-a^4}\cot^2(\theta/2)e^{2i\psi},
\label{Phipolar}
\end{equation}
and
\begin{equation}
A_\mu=\frac{e^{i\psi}}{(r^4-a^4)^{1/2}}
\left(\frac{2r^3\cot(\theta/2)}{a^4-r^4},-\frac12\cot^2(\theta/2),
i\cot(\theta/2),-\frac{i}{2}\sin(\theta)\right).
\label{Amupolar}
\end{equation}
We may also calculate the gauge-invariant field strength tensor,
\begin{equation}
F_{\mu\nu}=\partial_{[\mu}A_{\nu]}=
\frac{e^{i\psi}}{\sqrt{r^4-a^4}}
\left(
\begin{array}{cccc}
0 & -\frac{r^3}{r^4-a^4} & 0 & \frac{i r^3\sin\theta}{r^4-a^4}\\
\frac{r^3}{r^4-a^4} & 0 & -\frac{i}{2} & -\frac{i\cos\theta}{2}\\
0 & \frac{i}{2} & 0 & \frac{\sin\theta}{2}\\
-\frac{ir^3\sin\theta}{r^4-a^4} & \frac{i\cos\theta}{2} & 
-\frac{\sin\theta}{2} & 0
\end{array}\right),
\label{Fmunu}
\end{equation}
%where $D_\mu$ is the covariant derivative in the curvilinear coordinate system obtained by transforming the flat space metric from the lightcone 
% coordinate system via eqs.~(\ref{lightcone},
% \ref{Polardef}, \ref{psitrans}). 
As a cross-check, we have explicitly
verified (in the final coordinate system) that eq.~(\ref{Fmunu})
satisfies the Maxwell equations
\begin{equation}
D_\mu\,F^{\mu\nu}=0,
\label{Maxwell2}
\end{equation}
where $D_\mu$ is the covariant derivative in the curvilinear coordinate system obtained by transforming the flat space metric from the lightcone 
coordinate system via eqs.~(\ref{lightcone}, \ref{Polardef}, \ref{psitrans}), 
as well as the appropriate self-duality condition
\begin{equation}
F_{\mu\nu}=-\frac{\sqrt{g}}{2}\epsilon_{\mu\nu\alpha\beta}
F^{\alpha\beta},
\label{selfdual2}
\end{equation}
where $g$ is the determinant of the metric. One may also verify that
\begin{equation}
F_{\mu\nu}\, F^{\mu\nu}=
F_{\mu\nu}\, \tilde{F}^{\mu\nu}=0,
\label{FF}
\end{equation}
where the second condition follows from the first due to
self-duality. The vanishing of the field strength contracted with
itself implies that the gauge field has zero action, as does the
gravity solution~\footnote{The reader might be concerned that $F^2=0$
  implies that $F=0$, but the complexification of the field strength
  induced by the coordinate transformation of eq.~(\ref{psitrans})
  allows non-trivial field strengths to have vanishing action.}.  The
field strength also has a singularity as $r\rightarrow a$, mirroring
the bolt singularity in the gravity theory. However, unlike in the
gravity case, where the bolt is a genuine singularity in the metric,
its gauge theory counterpart is entirely spurious, having been
introduced by the coordinate transformation of
eq.~(\ref{psitrans}). Without this transformation, the gauge field is
linear in $a$ (and thus well-behaved). We stress again that it lives
in flat space.  \\

\begin{figure}
\begin{center}
\scalebox{0.6}{\includegraphics{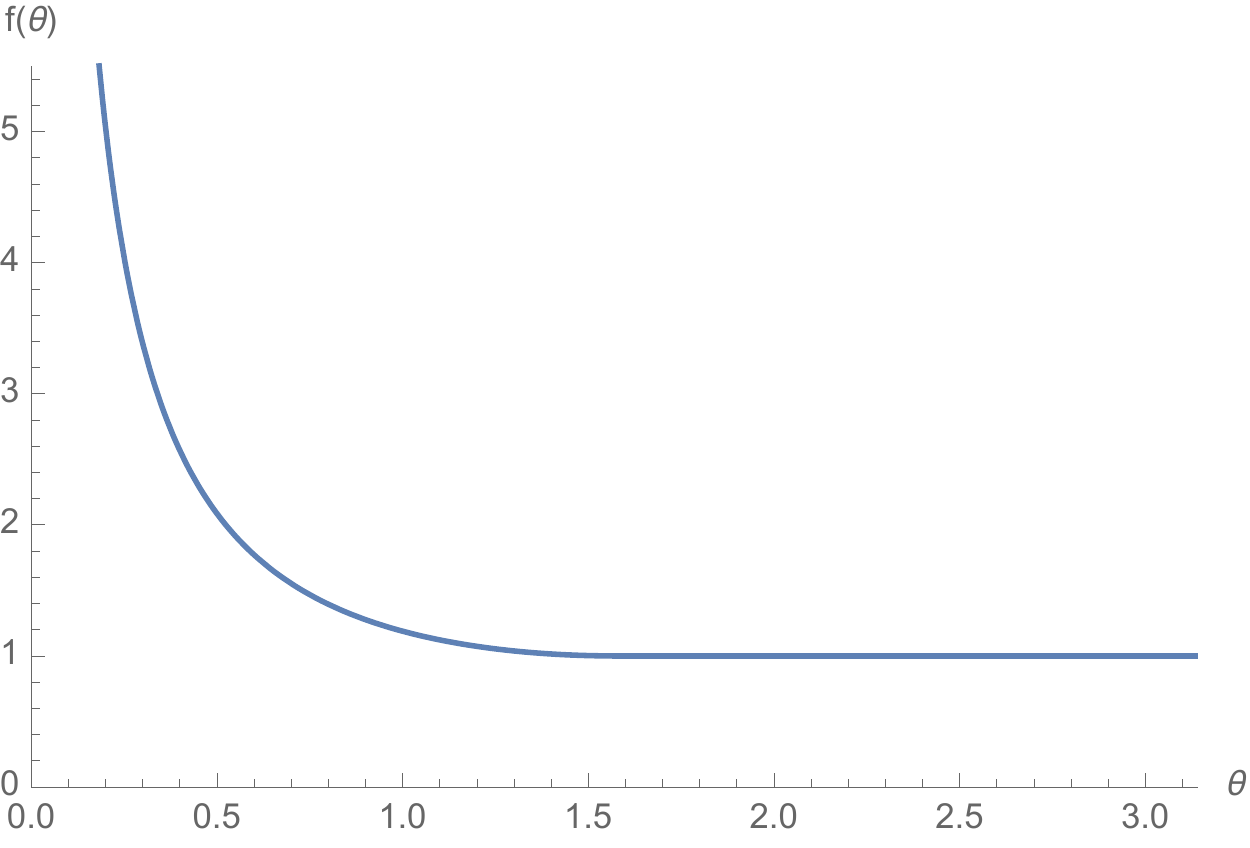}}
\caption{The function of eq.~(\ref{fdef}) entering the gauge
  transformation of eqs.~(\ref{AmuN}, \ref{alphadef}).}
\label{fig:fplot}
\end{center}
\end{figure}

\begin{figure}
\begin{center}
\scalebox{0.6}{\includegraphics{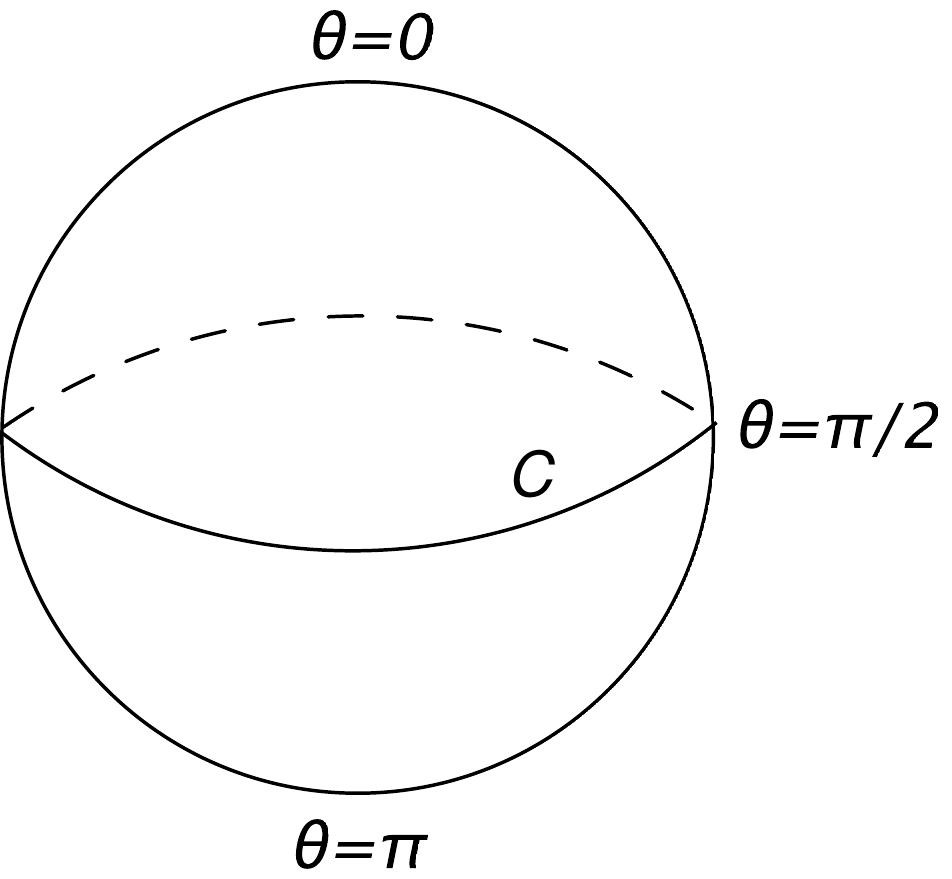}}
\caption{Northern and southern hemispheres for fixed $r$ and $\psi$,
  and the equator $C$.}
\label{fig:hemispheres}
\end{center}
\end{figure}

A noteworthy feature of the single copy field (in the polar
coordinates) is the presence of an apparent singularity as
$\theta\rightarrow 0$, which at first glance looks like the so-called
Dirac string singularity of a magnetic monopole~\cite{Dirac:1931kp},
associated with non-zero magnetic charge. However, we can completely
remove this singularity using a (globally defined) gauge
transformation
\begin{equation}
A_\mu\rightarrow 
A'_\mu\equiv A_\mu(r,\theta,\psi,\phi)-\partial_\mu \alpha(r,\theta,\psi,\phi),
\label{AmuN}
\end{equation}
where 
\begin{equation}
\alpha=\frac{2e^{i\psi}}{\sqrt{r^4-a^4}}f(\theta),
\label{alphadef}
\end{equation}
and
\begin{equation}
f(\theta)=\begin{cases}
\csc\theta,\quad \theta\leq\frac{\pi}{2};\\
1,\quad \theta>\frac{\pi}{2}.
\end{cases}
\label{fdef}
\end{equation}
A plot of this function is shown in figure~\ref{fig:fplot}. Both the
function and its first derivative are continuous at $\theta=\pi/2$.
One then has
\begin{align}
A'_\mu&=\begin{cases}\frac{e^{i\psi}}{\sqrt{r^4-a^4}}\left(
\frac{2r^3\tan(\theta/2)}{r^4-a^4},-\frac12\tan^2(\theta/2),
-i\tan(\theta/2),-\frac{i}{2}\sin\theta\right),\quad \theta\leq\pi/2;\\
\frac{e^{i\psi}}{\sqrt{r^4-a^4}}\left(
\frac{2r^3(2-\cot(\theta/2))}{r^4-a^4},-\frac12\cot^2(\theta/2),
-i(2-\cot(\theta/2)),-\frac{i}{2}\sin\theta\right),\quad \theta>\pi/2.
\end{cases}
\label{Amutrans}
\end{align}
It is easily checked that $A'_\mu$ is continuous at $\theta=\pi/2$,
and non-singular for all $\theta\in[0,\pi]$. The fact that the
string-like singularity has been removed suggests that there is no
magnetic charge. To verify this explicity, consider a surface at fixed
$\psi$ (which plays the role of the time coordinate), and $r$. This is
a sphere defined by spherical polar coordinates $(\theta,\phi)$, as
shown in figure~\ref{fig:hemispheres}. 
The magnetic charge is associated with the {\it first Chern number}
\begin{equation}
c_1=\frac{1}{4\pi}\int_{\Sigma} F_{\mu\nu}d\Sigma^{\mu\nu},
\label{Chern1}
\end{equation}
where the closed 2-surface $\Sigma$ is the above sphere, and we may
choose $\psi=0$ without loss of generality. We may carry out the
surface integral by splitting it into two pieces, for the northern and
southern hemispheres respectively. Using Stoke's theorem, we then have
\begin{align}
c_1&=\int_C dx^\mu \delta A'_\mu =0,
\label{c1calc}
\end{align}
where $\delta A'_\mu$ is the difference in the (transformed) gauge
field in the northern and southern hemispheres, and the line integral
is around the equator $C$. The second equality follows given that
$\delta A'_\mu=0$ on $C$.\\

We should not be surprised that the magnetic charge is zero. Taking
$\psi$ as a (periodic) time coordinate, we see from eq.~(\ref{Fmunu})
that the electric field points tangential to the $S^2$ surface spanned
by ($\theta$,$\phi$). Thus, there is no electric flux through a closed
surface enclosing the single copy solution, implying (via Gauss' Law)
that the electric charge is zero. Self-duality then dictates that the
magnetic charge must also be zero. This is also consistent with the
fact that field-strength falls off in a dipole-like way $\sim r^{-3}$
(for fixed $\psi$) at large distances: a dipole has no net charge. \\

We have seen that the single copy gauge field has vanishing action,
and the natural question arises of whether it can be given an
instanton interpretation, analogous to that of its gravitational
counterpart. However, there can be no abelian-like instanton
solutions, owing to the fact that one requires a solution to be pure
gauge at infinity, such that there is a topologically nontrivial map
from the boundary of spacetime ($S^3$) to the gauge group. There are
no such maps~\footnote{Put more formally, the third homotopy group
  $\pi_3[U(1)]$ is trivial.} to the abelian group
$U(1)$. Another way to see the absence of instanton solutions is to
note that they can be labelled by the second Chern number
\begin{equation}
c_2=-\frac{1}{8\pi^2}\int_{{\mathbb R}^4}{\rm Tr}[F\wedge F]
=-\frac{1}{8\pi^2}\int_{\partial {\mathbb R}^4\equiv S^3}
{\rm Tr}[A\wedge F],
\label{Chern}
\end{equation}
where we use differential form notation, such that $F=dA$ is the field
strength. The integral on the right-hand side vanishes, owing to the
steep fall-off of both $A$ and $F$ that can be surmised from
eqs.~(\ref{Amupolar}, \ref{Fmunu}, \ref{Amutrans}). \\

The vanishing of the first and second Chern numbers mean that there is
no non-trivial topological character to the single copy gauge field,
which can be related directly to the gravity solution. We saw that the
Eguchi-Hanson spacetime is topologically trivial before restricting
the period of $\psi$ as in eq.~(\ref{psirange}). Here we have taken
the single copy without imposing any restrictions on the background
(flat) spacetime that the gauge theory lives in. Thus, the topological
triviality of the single copy is entirely consistent with the similar
behaviour of the (unrestricted) gravity solution. We may then ask
whether it is possible to obtain topologically nontrivial behaviour by
restricting the range of $\psi$ in the electromagnetic
theory. However, the above arguments are unmodified by such a
restriction: $\psi$ plays the role of a time coordinate, and the lack
of magnetic charge at one time persists at later times. Furthermore,
the obstruction to obtaining a nontrivial second Chern number is
unchanged, in that the fields continue to fall off sharply at the
boundary of the spacetime. Thus, there seems no way to obtain a gauge
theory counterpart to the non-trivial topology of the gravity solution
after removing the bolt singularity in the latter. We stress again
that this is not surprising: non-trivial topology in the gravity
theory arises upon removing the bolt singularity. The latter has no
meaning in the gauge theory i.e. the apparent singularity at $r=a$ can
be removed by a coordinate transformation, which is {\it not} the same
as a gauge transformation. There is therefore no non-trivial
topological character to the gauge field. \\

\begin{figure}
\centering
\begin{subfigure}{.5\textwidth}
  \centering
  \scalebox{0.5}{\includegraphics{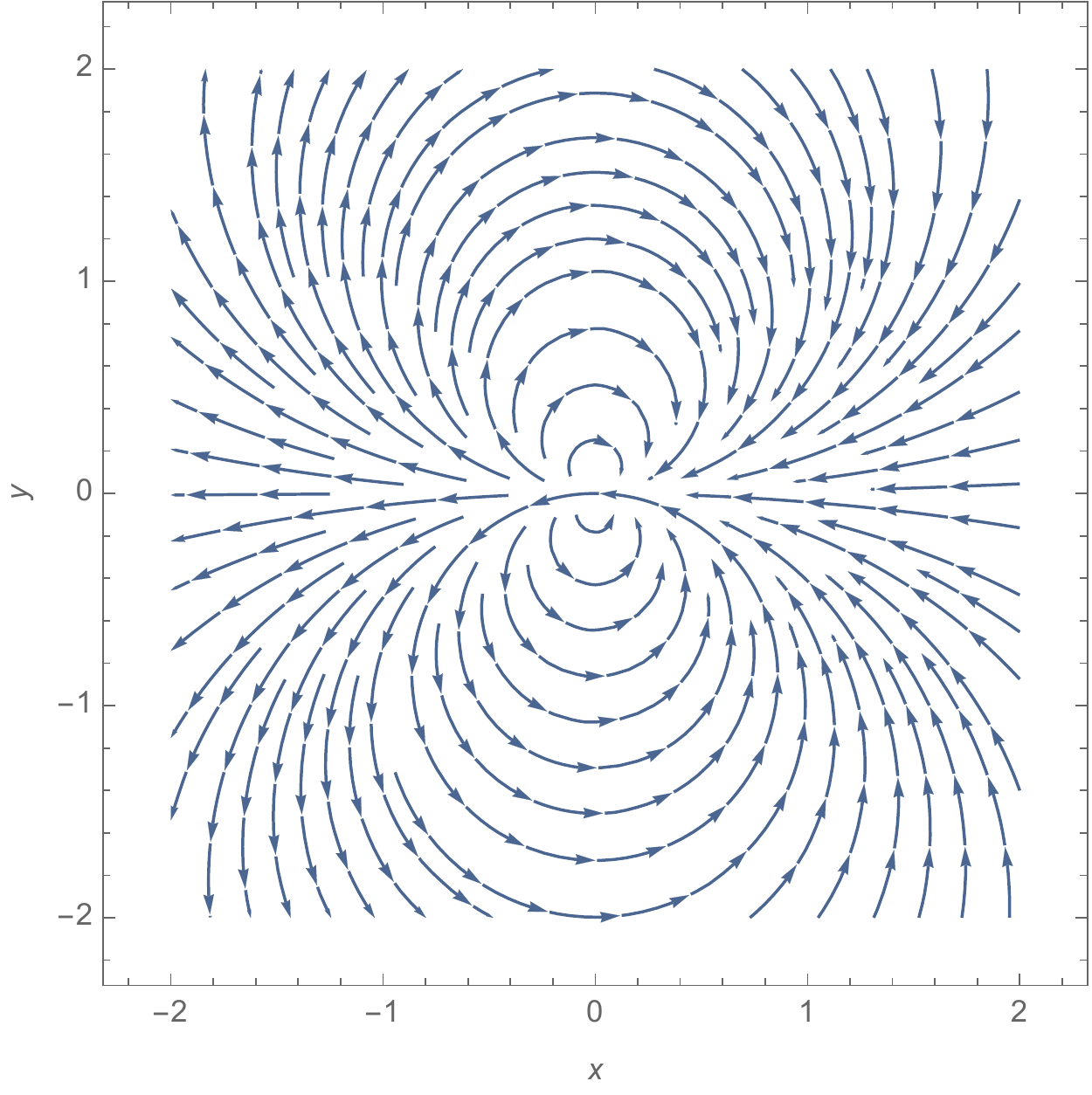}}
  \caption{}
  \label{fig:electric1a}
\end{subfigure}%
\begin{subfigure}{.5\textwidth}
  \centering
  \scalebox{0.5}{\includegraphics{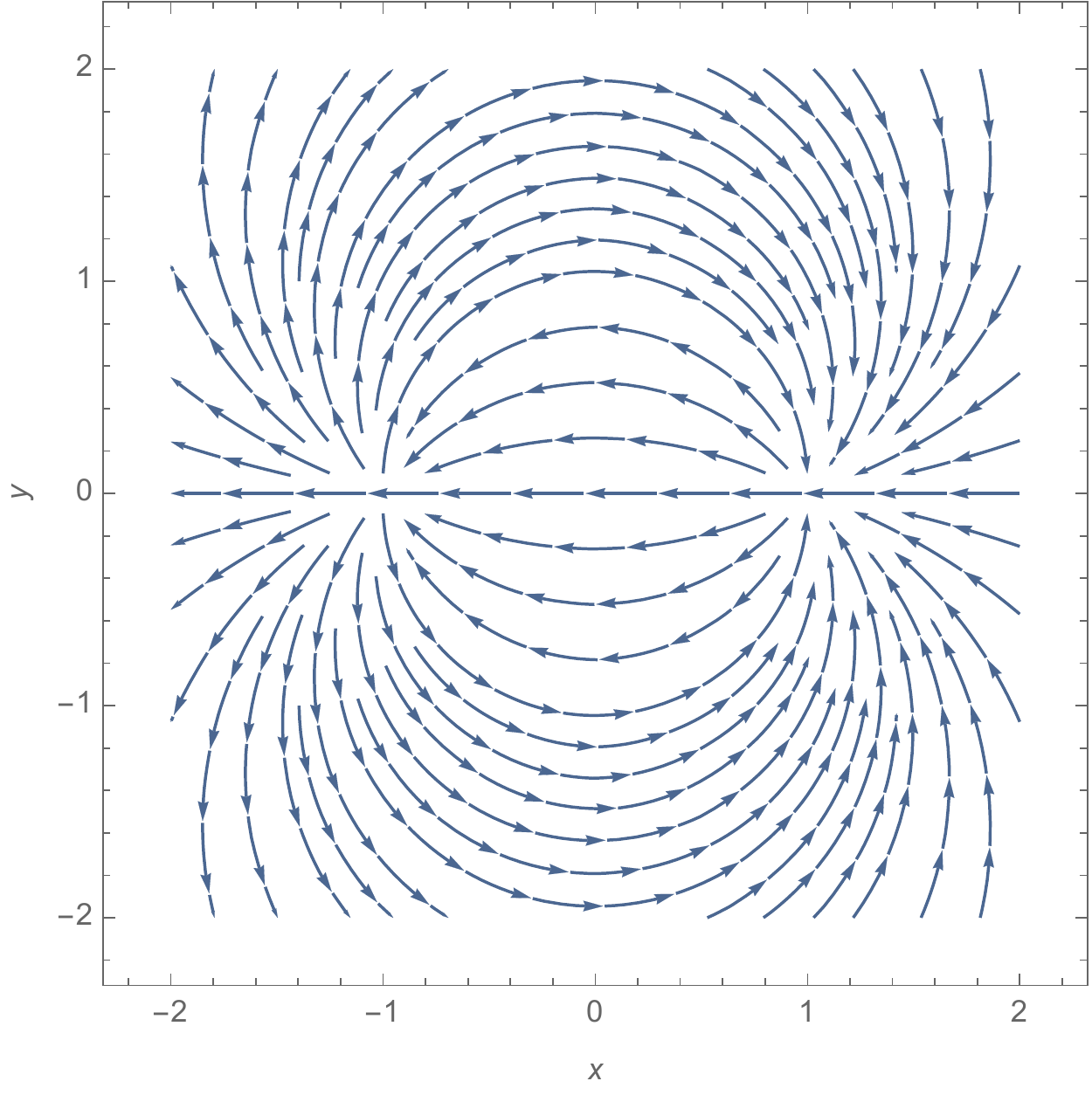}}
  \caption{}
  \label{fig:electric1b}
\end{subfigure}
\caption{The real part of the electric field of the Eguchi-Hanson
  single copy, using Cartesian coordinates in Minkowski signature: (a)
  for $(t,z)=(0,0)$; (b) for $(t,z)=(1,0)$. The imaginary part can be
  obtained by rotating the figures by $90^\circ$.}
\label{fig:electric1}
\end{figure}

\begin{figure}
\centering
\begin{subfigure}{.5\textwidth}
  \centering
  \scalebox{0.5}{\includegraphics{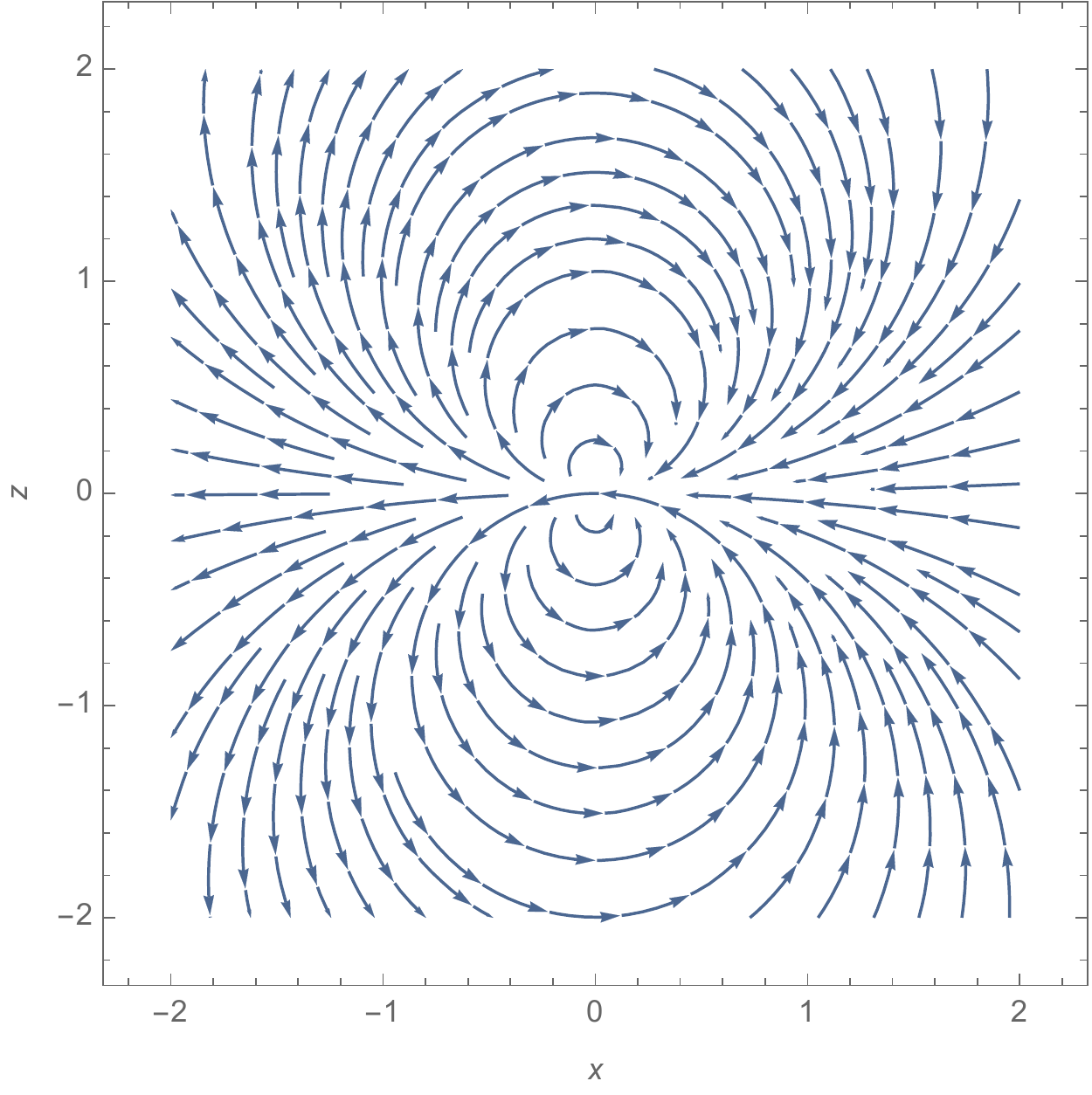}}
  \caption{}
  \label{fig:electric2a}
\end{subfigure}%
\begin{subfigure}{.5\textwidth}
  \centering
  \scalebox{0.5}{\includegraphics{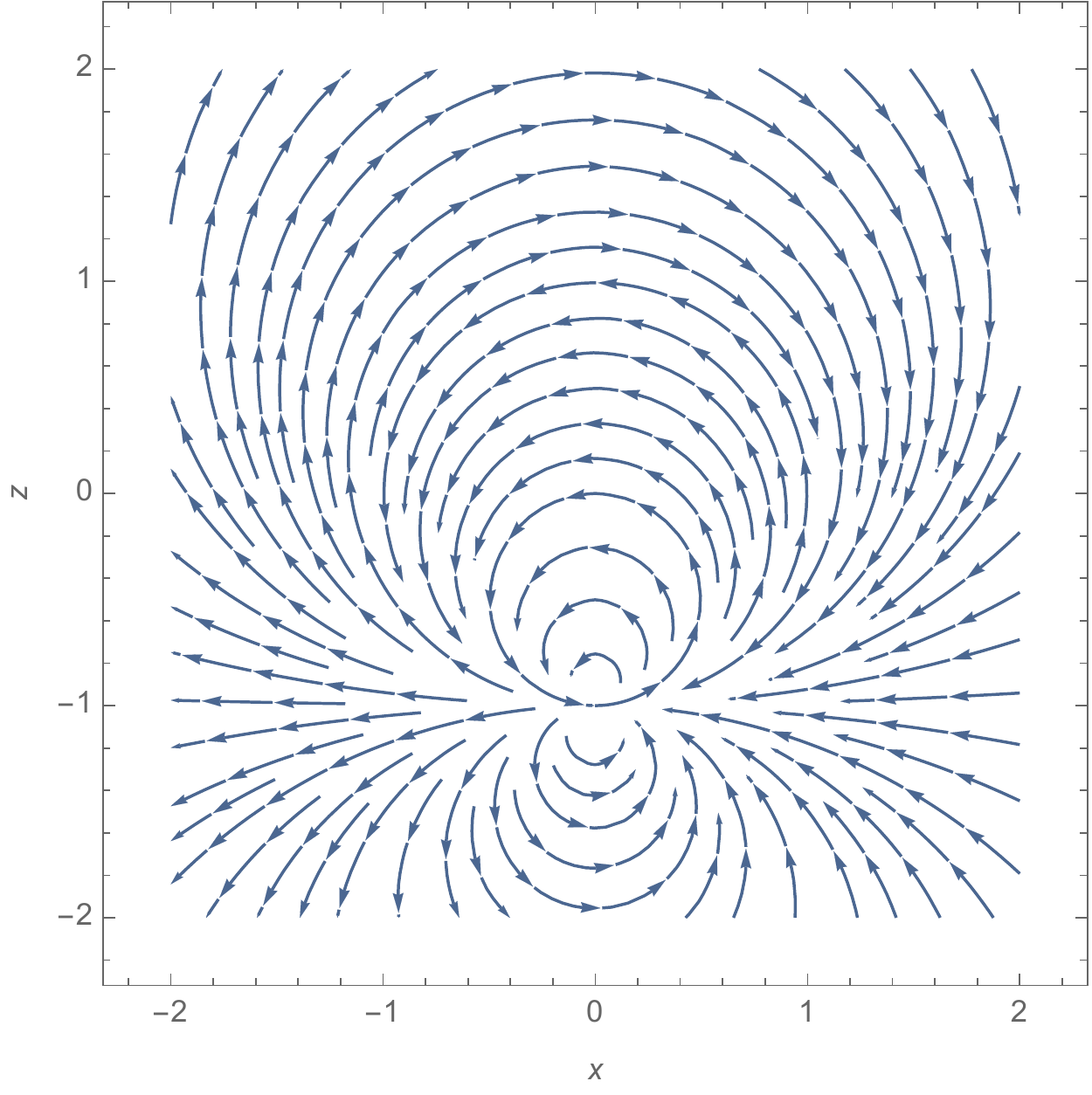}}
  \caption{}
  \label{fig:electric2b}
\end{subfigure}
\caption{The real part of the electric field of the Eguchi-Hanson
  single copy, using Cartesian coordinates in Minkowski signature: (a)
  for $(t,y)=(0,0)$; (b) for $(t,y)=(1,0)$. The imaginary part is
  zero.}
\label{fig:electric2}
\end{figure}

In order to corroborate the above discussion, it is instructive to
plot the electric field (related to the magnetic field by
self-duality) obtained from the single copy gauge field. To this end,
we use Cartesian coordinates in Minkowski signature, obtained from the
coordinates $(t,x,y,z)$ in eq.~(\ref{lightcone}) by sending
$t\rightarrow it$. In figure~\ref{fig:electric1}, we show the electric
field in the plane $z=0$, for two different times. 
At time $t=0$, the field indeed appears purely dipole-like. As time
increases, a disturbance propogates outwards at the speed of light,
such that the field is singular on the circular contour
$x^2+y^2=1$. Outside this region, however, the dipole-like form
remains. The field in other planes is more complicated. As an example,
we show the field in the $(x,z)$ plane in
figure~\ref{fig:electric2}. Again there is a lightcone singularity at
unit radius in the plane for $t=1$, but now the centre of the
dipole-like behaviour is progressively shifted as time progresses.
In all of the field configurations shown, there is clearly no net
electric charge and thus, by self-duality, no magnetic charge.

\section{Conclusion}
\label{sec:conclude}

In this paper, we have examined the double copy for exact classical
solutions of gauge theory and gravity that are Kerr-Schild, but also
self-dual. For this special class, one may consider a harmonic
function in light-cone coordinates, and interpret the Kerr-Schild
vector as a certain differential operator $\hat{k}_\mu$, which acts on
the harmonic function to produce gauge and graviton fields. The
harmonic function and gauge field are then the zeroth and single copy
of the graviton respectively.\\

As an example of this approach, we have constructed a single copy of
the well-known Eguchi-Hanson solution, which has been widely studied
over the years as a gravitational instanton. The Eguchi-Hanson
instanton has non-trivial topological structure, after removal of the
so-called bolt singularity occuring in the metric. Before removal of
this singularity however, the Euler characteristic and Hirzebruch
signature of the spacetime are both zero, implying trivial
topology. Consistent with this, we find that the single copy gauge
field is also topologically trivial, as it must be given that it is
abelian-like, and thus cannot exhibit nonzero instanton number if the
boundary of spacetime is $S^3$. The magnetic charge is also zero,
matching the fact that the field has dipole-like behaviour.\\

Our results are interesting in that one might have expected the single
copy of the Eguchi-Hanson metric to be a non-abelian instanton, such
as the well-known BPST instanton of
ref.~\cite{Belavin:1975fg}. Indeed, it was the discovery of the latter
solution that motivated the original study that led to the
Eguchi-Hanson metric~\cite{Eguchi:1978xp}. We find instead a single
copy field that is abelian-like, and has trivial topology, and it is
not the first time that the single copy of a well-known gravity
solution has turned out to be simpler than expected. Before
ref.~\cite{Monteiro:2014cda}, it was expected by many (including one
of the present authors) that the single copy of the Schwarzschild
solution would be a non-abelian object e.g. a monopole. However, the
single copy turns out to be a simple Coulomb charge. This was later
made sense of by studying the Taub-NUT solution~\cite{Luna:2015paa},
which confirmed that the single copy maps mass and NUT charge in the
gravity theory to electric and magnetic charge, respectively, in the
gauge theory. For Eguchi-Hanson, there is zero NUT charge and mass,
and thus we indeed expect no electric or magnetic charge in its single
copy.\\

Our results will also be useful in guiding studies of nonperturbative
aspects of the double copy. Currently, all known examples of the copy
involve objects cast in perturbation theory, or involving positive
powers of the coupling constant. It is not known whether the double
copy is a truly nonperturbative statement, expressing a much deeper
relationship between gauge and gravity theories than previously
thought. Furthermore, it is not even clear how to systematically
formulate such a relationship. One way forward might be to catalogue
known nonperturbative (or strong coupling) solutions between different
theories, before matching them up. To this end, some strong coupling
solutions of biadjoint scalar theory have been recently derived in
refs.~\cite{White:2016jzc,DeSmet:2017rve}.\\

Another way forward might be to use symmetries to characterise
nonperturbative behaviour on both sides of the double copy
correspondence, and to match up solutions according to
e.g. topological information. Here, we have found that the creation of
topological charge in the gravity theory does not necessarily have a
unique counterpart in the gauge theory. The removal of the bolt
singularity in the Eguchi-Hanson metric can be seen as a boundary
condition, which has no immediate analogue in gauge theory. \\

There remains the possibility that there may yet be topologically
nontrivial non-abelian solutions that map to Eguchi-Hanson, in
addition to the abelian-like gauge field considered here. The fact
that both an abelian and non-abelian quantity may map to the same
gravitational quantity is possible given that colour structure is
stripped off upon performing the double copy, and indeed such
behaviour has been seen before in the study of infrared
singularities~\cite{Oxburgh:2012zr}. One way to proceed may be to
embed the abelian-like field seen here in a fully non-abelian setting
using the well-known t'Hooft ansatz~\cite{tHooft:1976snw}.\\

In summary, the study of classical aspects of the double copy,
including possible nonperturbative aspects, is in its infancy, with
many exciting avenues still to be explored. We hope that our present
study sheds some light on some of these possibilities.

\section*{Acknowledgments}

We thank Joe Hayling for remarks that initiated this study, as well as
Andreas Brandhuber, Gabriele Travaglini and Costis Papageorgakis for
numerous illuminating chats. We are also grateful to Donal O'Connell
and Ricardo Monteiro for discussions and collaboration on related
topics. CDW and DB are supported by the UK Science and Technology
Facilities Council (STFC). AL is supported in part by the Department
of Energy under Award Number DESC000993.

\bibliography{refs.bib}

\providecommand{\href}[2]{#2}\begingroup\raggedright\begin{thebibliography}{10}

\bibitem{Bern:2008qj}
Z.~Bern, J.~Carrasco, and H.~Johansson, ``{New Relations for Gauge-Theory
  Amplitudes},'' {\em Phys.Rev.} {\bf D78} (2008) 085011,
\href{http://www.arXiv.org/abs/0805.3993}{{\tt 0805.3993}}.
%%CITATION = ARXIV:0805.3993;%%.

\bibitem{Bern:2010ue}
Z.~Bern, J.~J.~M. Carrasco, and H.~Johansson, ``{Perturbative Quantum Gravity
  as a Double Copy of Gauge Theory},'' {\em Phys.Rev.Lett.} {\bf 105} (2010)
  061602, \href{http://www.arXiv.org/abs/1004.0476}{{\tt 1004.0476}}.

\bibitem{Bern:2010yg}
Z.~Bern, T.~Dennen, Y.-t. Huang, and M.~Kiermaier, ``{Gravity as the Square of
  Gauge Theory},'' {\em Phys.Rev.} {\bf D82} (2010) 065003,
  \href{http://www.arXiv.org/abs/1004.0693}{{\tt 1004.0693}}.

\bibitem{BjerrumBohr:2009rd}
N.~Bjerrum-Bohr, P.~H. Damgaard, and P.~Vanhove, ``{Minimal Basis for Gauge
  Theory Amplitudes},'' {\em Phys.Rev.Lett.} {\bf 103} (2009) 161602,
\href{http://www.arXiv.org/abs/0907.1425}{{\tt 0907.1425}}.
%%CITATION = ARXIV:0907.1425;%%.

\bibitem{Stieberger:2009hq}
S.~Stieberger, ``{Open and Closed vs. Pure Open String Disk Amplitudes},''
\href{http://www.arXiv.org/abs/0907.2211}{{\tt 0907.2211}}.
%%CITATION = ARXIV:0907.2211;%%.

\bibitem{BjerrumBohr:2010zs}
N.~Bjerrum-Bohr, P.~H. Damgaard, T.~Sondergaard, and P.~Vanhove, ``{Monodromy
  and Jacobi-like Relations for Color-Ordered Amplitudes},'' {\em JHEP} {\bf
  1006} (2010) 003,
\href{http://www.arXiv.org/abs/1003.2403}{{\tt 1003.2403}}.
%%CITATION = ARXIV:1003.2403;%%.

\bibitem{Feng:2010my}
B.~Feng, R.~Huang, and Y.~Jia, ``{Gauge Amplitude Identities by On-shell
  Recursion Relation in S-matrix Program},'' {\em Phys.Lett.} {\bf B695} (2011)
  350--353,
\href{http://www.arXiv.org/abs/1004.3417}{{\tt 1004.3417}}.
%%CITATION = ARXIV:1004.3417;%%.

\bibitem{Tye:2010dd}
S.~Henry~Tye and Y.~Zhang, ``{Dual Identities inside the Gluon and the Graviton
  Scattering Amplitudes},'' {\em JHEP} {\bf 1006} (2010) 071,
\href{http://www.arXiv.org/abs/1003.1732}{{\tt 1003.1732}}.
%%CITATION = ARXIV:1003.1732;%%.

\bibitem{Mafra:2011kj}
C.~R. Mafra, O.~Schlotterer, and S.~Stieberger, ``{Explicit BCJ Numerators from
  Pure Spinors},'' {\em JHEP} {\bf 1107} (2011) 092,
\href{http://www.arXiv.org/abs/1104.5224}{{\tt 1104.5224}}.
%%CITATION = ARXIV:1104.5224;%%.

\bibitem{Monteiro:2011pc}
R.~Monteiro and D.~O'Connell, ``{The Kinematic Algebra From the Self-Dual
  Sector},'' {\em JHEP} {\bf 1107} (2011) 007,
\href{http://www.arXiv.org/abs/1105.2565}{{\tt 1105.2565}}.
%%CITATION = ARXIV:1105.2565;%%.

\bibitem{BjerrumBohr:2012mg}
N.~Bjerrum-Bohr, P.~H. Damgaard, R.~Monteiro, and D.~O'Connell, ``{Algebras for
  Amplitudes},'' {\em JHEP} {\bf 1206} (2012) 061,
\href{http://www.arXiv.org/abs/1203.0944}{{\tt 1203.0944}}.
%%CITATION = ARXIV:1203.0944;%%.

\bibitem{Kawai:1985xq}
H.~Kawai, D.~Lewellen, and S.~Tye, ``{A Relation Between Tree Amplitudes of
  Closed and Open Strings},'' {\em Nucl.Phys.} {\bf B269} (1986)
1.
%%CITATION = NUPHA,B269,1;%%.

\bibitem{Bern:1998ug}
Z.~Bern, L.~J. Dixon, D.~Dunbar, M.~Perelstein, and J.~Rozowsky, ``{On the
  relationship between Yang-Mills theory and gravity and its implication for
  ultraviolet divergences},'' {\em Nucl.Phys.} {\bf B530} (1998) 401--456,
\href{http://www.arXiv.org/abs/hep-th/9802162}{{\tt hep-th/9802162}}.
%%CITATION = HEP-TH/9802162;%%.

\bibitem{Green:1982sw}
M.~B. Green, J.~H. Schwarz, and L.~Brink, ``{N=4 Yang-Mills and N=8
  Supergravity as Limits of String Theories},'' {\em Nucl.Phys.} {\bf B198}
  (1982)
474--492.
%%CITATION = NUPHA,B198,474;%%.

\bibitem{Bern:1997nh}
Z.~Bern, J.~Rozowsky, and B.~Yan, ``{Two loop four gluon amplitudes in N=4
  superYang-Mills},'' {\em Phys.Lett.} {\bf B401} (1997) 273--282,
\href{http://www.arXiv.org/abs/hep-ph/9702424}{{\tt hep-ph/9702424}}.
%%CITATION = HEP-PH/9702424;%%.

\bibitem{Carrasco:2011mn}
J.~J. Carrasco and H.~Johansson, ``{Five-Point Amplitudes in N=4
  Super-Yang-Mills Theory and N=8 Supergravity},'' {\em Phys.Rev.} {\bf D85}
  (2012) 025006,
\href{http://www.arXiv.org/abs/1106.4711}{{\tt 1106.4711}}.
%%CITATION = ARXIV:1106.4711;%%.

\bibitem{Carrasco:2012ca}
J.~J.~M. Carrasco, M.~Chiodaroli, M.~Günaydin, and R.~Roiban, ``{One-loop
  four-point amplitudes in pure and matter-coupled N=4 supergravity},'' {\em
  JHEP} {\bf 1303} (2013) 056,
\href{http://www.arXiv.org/abs/1212.1146}{{\tt 1212.1146}}.
%%CITATION = ARXIV:1212.1146;%%.

\bibitem{Mafra:2012kh}
C.~R. Mafra and O.~Schlotterer, ``{The Structure of n-Point One-Loop Open
  Superstring Amplitudes},'' {\em JHEP} {\bf 1408} (2014) 099,
\href{http://www.arXiv.org/abs/1203.6215}{{\tt 1203.6215}}.
%%CITATION = ARXIV:1203.6215;%%.

\bibitem{Boels:2013bi}
R.~H. Boels, R.~S. Isermann, R.~Monteiro, and D.~O'Connell,
  ``{Colour-Kinematics Duality for One-Loop Rational Amplitudes},'' {\em JHEP}
  {\bf 1304} (2013) 107,
\href{http://www.arXiv.org/abs/1301.4165}{{\tt 1301.4165}}.
%%CITATION = ARXIV:1301.4165;%%.

\bibitem{Bjerrum-Bohr:2013iza}
N.~E.~J. Bjerrum-Bohr, T.~Dennen, R.~Monteiro, and D.~O'Connell, ``{Integrand
  Oxidation and One-Loop Colour-Dual Numerators in N=4 Gauge Theory},'' {\em
  JHEP} {\bf 1307} (2013) 092,
\href{http://www.arXiv.org/abs/1303.2913}{{\tt 1303.2913}}.
%%CITATION = ARXIV:1303.2913;%%.

\bibitem{Bern:2013yya}
Z.~Bern, S.~Davies, T.~Dennen, Y.-t. Huang, and J.~Nohle, ``{Color-Kinematics
  Duality for Pure Yang-Mills and Gravity at One and Two Loops},''
\href{http://www.arXiv.org/abs/1303.6605}{{\tt 1303.6605}}.
%%CITATION = ARXIV:1303.6605;%%.

\bibitem{Bern:2013qca}
Z.~Bern, S.~Davies, and T.~Dennen, ``{The Ultraviolet Structure of Half-Maximal
  Supergravity with Matter Multiplets at Two and Three Loops},'' {\em
  Phys.Rev.} {\bf D88} (2013) 065007,
\href{http://www.arXiv.org/abs/1305.4876}{{\tt 1305.4876}}.
%%CITATION = ARXIV:1305.4876;%%.

\bibitem{Nohle:2013bfa}
J.~Nohle, ``{Color-Kinematics Duality in One-Loop Four-Gluon Amplitudes with
  Matter},''
\href{http://www.arXiv.org/abs/1309.7416}{{\tt 1309.7416}}.
%%CITATION = ARXIV:1309.7416;%%.

\bibitem{Bern:2013uka}
Z.~Bern, S.~Davies, T.~Dennen, A.~V. Smirnov, and V.~A. Smirnov, ``{Ultraviolet
  Properties of N=4 Supergravity at Four Loops},'' {\em Phys.Rev.Lett.} {\bf
  111} (2013), no.~23, 231302,
\href{http://www.arXiv.org/abs/1309.2498}{{\tt 1309.2498}}.
%%CITATION = ARXIV:1309.2498;%%.

\bibitem{Naculich:2013xa}
S.~G. Naculich, H.~Nastase, and H.~J. Schnitzer, ``{All-loop infrared-divergent
  behavior of most-subleading-color gauge-theory amplitudes},'' {\em JHEP} {\bf
  1304} (2013) 114,
\href{http://www.arXiv.org/abs/1301.2234}{{\tt 1301.2234}}.
%%CITATION = ARXIV:1301.2234;%%.

\bibitem{Du:2014uua}
Y.-J. Du, B.~Feng, and C.-H. Fu, ``{Dual-color decompositions at one-loop level
  in Yang-Mills theory},''
\href{http://www.arXiv.org/abs/1402.6805}{{\tt 1402.6805}}.
%%CITATION = ARXIV:1402.6805;%%.

\bibitem{Mafra:2014gja}
C.~R. Mafra and O.~Schlotterer, ``{Towards one-loop SYM amplitudes from the
  pure spinor BRST cohomology},'' {\em Fortsch.Phys.} {\bf 63} (2015), no.~2,
  105--131,
\href{http://www.arXiv.org/abs/1410.0668}{{\tt 1410.0668}}.
%%CITATION = ARXIV:1410.0668;%%.

\bibitem{Bern:2014sna}
Z.~Bern, S.~Davies, and T.~Dennen, ``{Enhanced Ultraviolet Cancellations in N =
  5 Supergravity at Four Loop},''
\href{http://www.arXiv.org/abs/1409.3089}{{\tt 1409.3089}}.
%%CITATION = ARXIV:1409.3089;%%.

\bibitem{Mafra:2015mja}
C.~R. Mafra and O.~Schlotterer, ``{Two-loop five-point amplitudes of super
  Yang-Mills and supergravity in pure spinor superspace},''
\href{http://www.arXiv.org/abs/1505.02746}{{\tt 1505.02746}}.
%%CITATION = ARXIV:1505.02746;%%.

\bibitem{He:2015wgf}
S.~He, R.~Monteiro, and O.~Schlotterer, ``{String-inspired BCJ numerators for
  one-loop MHV amplitudes},'' {\em JHEP} {\bf 01} (2016) 171,
\href{http://www.arXiv.org/abs/1507.06288}{{\tt 1507.06288}}.
%%CITATION = ARXIV:1507.06288;%%.

\bibitem{Bern:2015ooa}
Z.~Bern, S.~Davies, and J.~Nohle, ``{Double-Copy Constructions and Unitarity
  Cuts},''
\href{http://www.arXiv.org/abs/1510.03448}{{\tt 1510.03448}}.
%%CITATION = ARXIV:1510.03448;%%.

\bibitem{Mogull:2015adi}
G.~Mogull and D.~O'Connell, ``{Overcoming Obstacles to Colour-Kinematics
  Duality at Two Loops},'' {\em JHEP} {\bf 12} (2015) 135,
\href{http://www.arXiv.org/abs/1511.06652}{{\tt 1511.06652}}.
%%CITATION = ARXIV:1511.06652;%%.

\bibitem{Chiodaroli:2015rdg}
M.~Chiodaroli, M.~Gunaydin, H.~Johansson, and R.~Roiban, ``{Spontaneously
  Broken Yang-Mills-Einstein Supergravities as Double Copies},''
\href{http://www.arXiv.org/abs/1511.01740}{{\tt 1511.01740}}.
%%CITATION = ARXIV:1511.01740;%%.

\bibitem{Bern:2017ucb}
Z.~Bern, J.~J.~M. Carrasco, W.-M. Chen, H.~Johansson, R.~Roiban, and M.~Zeng,
  ``{The Five-Loop Four-Point Integrand of N=8 Supergravity as a Generalized
  Double Copy},''
\href{http://www.arXiv.org/abs/1708.06807}{{\tt 1708.06807}}.
%%CITATION = ARXIV:1708.06807;%%.

\bibitem{Johansson:2015oia}
H.~Johansson and A.~Ochirov, ``{Color-Kinematics Duality for QCD Amplitudes},''
  {\em JHEP} {\bf 01} (2016) 170,
\href{http://www.arXiv.org/abs/1507.00332}{{\tt 1507.00332}}.
%%CITATION = ARXIV:1507.00332;%%.

\bibitem{Oxburgh:2012zr}
S.~Oxburgh and C.~White, ``{BCJ duality and the double copy in the soft
  limit},'' {\em JHEP} {\bf 1302} (2013) 127,
\href{http://www.arXiv.org/abs/1210.1110}{{\tt 1210.1110}}.
%%CITATION = ARXIV:1210.1110;%%.

\bibitem{White:2011yy}
C.~D. White, ``{Factorization Properties of Soft Graviton Amplitudes},'' {\em
  JHEP} {\bf 1105} (2011) 060, \href{http://www.arXiv.org/abs/1103.2981}{{\tt
  1103.2981}}.

\bibitem{Melville:2013qca}
S.~Melville, S.~Naculich, H.~Schnitzer, and C.~White, ``{Wilson line approach
  to gravity in the high energy limit},'' {\em Phys.Rev.} {\bf D89} (2014)
  025009,
\href{http://www.arXiv.org/abs/1306.6019}{{\tt 1306.6019}}.
%%CITATION = ARXIV:1306.6019;%%.

\bibitem{Luna:2016idw}
A.~Luna, S.~Melville, S.~G. Naculich, and C.~D. White, ``{Next-to-soft
  corrections to high energy scattering in QCD and gravity},'' {\em JHEP} {\bf
  01} (2017) 052,
\href{http://www.arXiv.org/abs/1611.02172}{{\tt 1611.02172}}.
%%CITATION = ARXIV:1611.02172;%%.

\bibitem{Saotome:2012vy}
R.~Saotome and R.~Akhoury, ``{Relationship Between Gravity and Gauge Scattering
  in the High Energy Limit},'' {\em JHEP} {\bf 1301} (2013) 123,
\href{http://www.arXiv.org/abs/1210.8111}{{\tt 1210.8111}}.
%%CITATION = ARXIV:1210.8111;%%.

\bibitem{Vera:2012ds}
A.~Sabio~Vera, E.~Serna~Campillo, and M.~A. Vazquez-Mozo, ``{Color-Kinematics
  Duality and the Regge Limit of Inelastic Amplitudes},'' {\em JHEP} {\bf 1304}
  (2013) 086,
\href{http://www.arXiv.org/abs/1212.5103}{{\tt 1212.5103}}.
%%CITATION = ARXIV:1212.5103;%%.

\bibitem{Johansson:2013nsa}
H.~Johansson, A.~Sabio~Vera, E.~Serna~Campillo, and M.~Ã. Vázquez-Mozo,
  ``{Color-Kinematics Duality in Multi-Regge Kinematics and Dimensional
  Reduction},'' {\em JHEP} {\bf 1310} (2013) 215,
\href{http://www.arXiv.org/abs/1307.3106}{{\tt 1307.3106}}.
%%CITATION = ARXIV:1307.3106;%%.

\bibitem{Johansson:2013aca}
H.~Johansson, A.~Sabio~Vera, E.~Serna~Campillo, and M.~A. Vazquez-Mozo,
  ``{Color-kinematics duality and dimensional reduction for graviton emission
  in Regge limit},''
\href{http://www.arXiv.org/abs/1310.1680}{{\tt 1310.1680}}.
%%CITATION = ARXIV:1310.1680;%%.

\bibitem{Monteiro:2014cda}
R.~Monteiro, D.~O'Connell, and C.~D. White, ``{Black holes and the double
  copy},'' {\em JHEP} {\bf 1412} (2014) 056,
\href{http://www.arXiv.org/abs/1410.0239}{{\tt 1410.0239}}.
%%CITATION = ARXIV:1410.0239;%%.

\bibitem{Luna:2015paa}
A.~Luna, R.~Monteiro, D.~O'Connell, and C.~D. White, ``{The classical double
  copy for Taub–NUT spacetime},'' {\em Phys. Lett.} {\bf B750} (2015)
  272--277,
\href{http://www.arXiv.org/abs/1507.01869}{{\tt 1507.01869}}.
%%CITATION = ARXIV:1507.01869;%%.

\bibitem{Luna:2016due}
A.~Luna, R.~Monteiro, I.~Nicholson, D.~O'Connell, and C.~D. White, ``{The
  double copy: Bremsstrahlung and accelerating black holes},''
\href{http://www.arXiv.org/abs/1603.05737}{{\tt 1603.05737}}.
%%CITATION = ARXIV:1603.05737;%%.

\bibitem{Goldberger:2016iau}
W.~D. Goldberger and A.~K. Ridgway, ``{Radiation and the classical double copy
  for color charges},'' {\em Phys. Rev.} {\bf D95} (2017), no.~12, 125010,
\href{http://www.arXiv.org/abs/1611.03493}{{\tt 1611.03493}}.
%%CITATION = ARXIV:1611.03493;%%.

\bibitem{Anastasiou:2014qba}
A.~Anastasiou, L.~Borsten, M.~J. Duff, L.~J. Hughes, and S.~Nagy, ``{Yang-Mills
  origin of gravitational symmetries},'' {\em Phys. Rev. Lett.} {\bf 113}
  (2014), no.~23, 231606,
\href{http://www.arXiv.org/abs/1408.4434}{{\tt 1408.4434}}.
%%CITATION = ARXIV:1408.4434;%%.

\bibitem{Borsten:2015pla}
L.~Borsten and M.~J. Duff, ``{Gravity as the square of Yang–Mills?},'' {\em
  Phys. Scripta} {\bf 90} (2015) 108012,
\href{http://www.arXiv.org/abs/1602.08267}{{\tt 1602.08267}}.
%%CITATION = ARXIV:1602.08267;%%.

\bibitem{Anastasiou:2016csv}
A.~Anastasiou, L.~Borsten, M.~J. Duff, M.~J. Hughes, A.~Marrani, S.~Nagy, and
  M.~Zoccali, ``{Twin supergravities from Yang-Mills theory squared},'' {\em
  Phys. Rev.} {\bf D96} (2017), no.~2, 026013,
\href{http://www.arXiv.org/abs/1610.07192}{{\tt 1610.07192}}.
%%CITATION = ARXIV:1610.07192;%%.

\bibitem{Anastasiou:2017nsz}
A.~Anastasiou, L.~Borsten, M.~J. Duff, A.~Marrani, S.~Nagy, and M.~Zoccali,
  ``{Are all supergravity theories Yang-Mills squared?},''
\href{http://www.arXiv.org/abs/1707.03234}{{\tt 1707.03234}}.
%%CITATION = ARXIV:1707.03234;%%.

\bibitem{Cardoso:2016ngt}
G.~L. Cardoso, S.~Nagy, and S.~Nampuri, ``{A double copy for $ \mathcal{N}=2 $
  supergravity: a linearised tale told on-shell},'' {\em JHEP} {\bf 10} (2016)
  127,
\href{http://www.arXiv.org/abs/1609.05022}{{\tt 1609.05022}}.
%%CITATION = ARXIV:1609.05022;%%.

\bibitem{Borsten:2017jpt}
L.~Borsten, ``{On $D=6$, $\mathcal{N}=(2,0)$ and $\mathcal{N}=(4,0)$
  theories},''
\href{http://www.arXiv.org/abs/1708.02573}{{\tt 1708.02573}}.
%%CITATION = ARXIV:1708.02573;%%.

\bibitem{Anastasiou:2017taf}
A.~Anastasiou, L.~Borsten, M.~J. Duff, A.~Marrani, S.~Nagy, and M.~Zoccali,
  ``{The Mile High Magic Pyramid},''
\newblock 2017.
\newblock
\href{http://www.arXiv.org/abs/1711.08476}{{\tt 1711.08476}}.
\newblock
%%CITATION = ARXIV:1711.08476;%%.

\bibitem{Anastasiou:2018rdx}
A.~Anastasiou, L.~Borsten, M.~J. Duff, S.~Nagy, and M.~Zoccali, ``{BRST
  squared},''
\href{http://www.arXiv.org/abs/1807.02486}{{\tt 1807.02486}}.
%%CITATION = ARXIV:1807.02486;%%.

\bibitem{LopesCardoso:2018xes}
G.~Lopes~Cardoso, G.~Inverso, S.~Nagy, and S.~Nampuri, ``{Comments on the
  double copy construction for gravitational theories},'' in {\em {17th
  Hellenic School and Workshops on Elementary Particle Physics and Gravity
  (CORFU2017) Corfu, Greece, September 2-28, 2017}}.
\newblock 2018.
\newblock
\href{http://www.arXiv.org/abs/1803.07670}{{\tt 1803.07670}}.
\newblock
%%CITATION = ARXIV:1803.07670;%%.

\bibitem{Goldberger:2017frp}
W.~D. Goldberger, S.~G. Prabhu, and J.~O. Thompson, ``{Classical gluon and
  graviton radiation from the bi-adjoint scalar double copy},''
\href{http://www.arXiv.org/abs/1705.09263}{{\tt 1705.09263}}.
%%CITATION = ARXIV:1705.09263;%%.

\bibitem{Goldberger:2017vcg}
W.~D. Goldberger and A.~K. Ridgway, ``{Bound states and the classical double
  copy},'' {\em Phys. Rev.} {\bf D97} (2018), no.~8, 085019,
\href{http://www.arXiv.org/abs/1711.09493}{{\tt 1711.09493}}.
%%CITATION = ARXIV:1711.09493;%%.

\bibitem{Goldberger:2017ogt}
W.~D. Goldberger, J.~Li, and S.~G. Prabhu, ``{Spinning particles, axion
  radiation, and the classical double copy},'' {\em Phys. Rev.} {\bf D97}
  (2018), no.~10, 105018,
\href{http://www.arXiv.org/abs/1712.09250}{{\tt 1712.09250}}.
%%CITATION = ARXIV:1712.09250;%%.

\bibitem{Luna:2016hge}
A.~Luna, R.~Monteiro, I.~Nicholson, A.~Ochirov, D.~O'Connell, N.~Westerberg,
  and C.~D. White, ``{Perturbative spacetimes from Yang-Mills theory},'' {\em
  JHEP} {\bf 04} (2017) 069,
\href{http://www.arXiv.org/abs/1611.07508}{{\tt 1611.07508}}.
%%CITATION = ARXIV:1611.07508;%%.

\bibitem{Luna:2017dtq}
A.~Luna, I.~Nicholson, D.~O'Connell, and C.~D. White, ``{Inelastic Black Hole
  Scattering from Charged Scalar Amplitudes},'' {\em JHEP} {\bf 03} (2018) 044,
\href{http://www.arXiv.org/abs/1711.03901}{{\tt 1711.03901}}.
%%CITATION = ARXIV:1711.03901;%%.

\bibitem{Shen:2018ebu}
C.-H. Shen, ``{Gravitational Radiation from Color-Kinematics Duality},''
\href{http://www.arXiv.org/abs/1806.07388}{{\tt 1806.07388}}.
%%CITATION = ARXIV:1806.07388;%%.

\bibitem{Levi:2018nxp}
M.~Levi, ``{Effective Field Theories of Post-Newtonian Gravity},''
\href{http://www.arXiv.org/abs/1807.01699}{{\tt 1807.01699}}.
%%CITATION = ARXIV:1807.01699;%%.

\bibitem{Plefka:2018dpa}
J.~Plefka, J.~Steinhoff, and W.~Wormsbecher, ``{Effective action of dilaton
  gravity as the classical double copy of Yang-Mills theory},''
\href{http://www.arXiv.org/abs/1807.09859}{{\tt 1807.09859}}.
%%CITATION = ARXIV:1807.09859;%%.

\bibitem{Cheung:2018wkq}
C.~Cheung, I.~Z. Rothstein, and M.~P. Solon, ``{From Scattering Amplitudes to
  Classical Potentials in the Post-Minkowskian Expansion},''
\href{http://www.arXiv.org/abs/1808.02489}{{\tt 1808.02489}}.
%%CITATION = ARXIV:1808.02489;%%.

\bibitem{Carrillo-Gonzalez:2018pjk}
M.~Carrillo-Gonzalez, R.~Penco, and M.~Trodden, ``{Radiation of scalar modes
  and the classical double copy},''
\href{http://www.arXiv.org/abs/1809.04611}{{\tt 1809.04611}}.
%%CITATION = ARXIV:1809.04611;%%.

\bibitem{Cachazo:2013hca}
F.~Cachazo, S.~He, and E.~Y. Yuan, ``{Scattering of Massless Particles in
  Arbitrary Dimensions},'' {\em Phys. Rev. Lett.} {\bf 113} (2014), no.~17,
  171601,
\href{http://www.arXiv.org/abs/1307.2199}{{\tt 1307.2199}}.
%%CITATION = ARXIV:1307.2199;%%.

\bibitem{Cachazo:2013iea}
F.~Cachazo, S.~He, and E.~Y. Yuan, ``{Scattering of Massless Particles:
  Scalars, Gluons and Gravitons},'' {\em JHEP} {\bf 07} (2014) 033,
\href{http://www.arXiv.org/abs/1309.0885}{{\tt 1309.0885}}.
%%CITATION = ARXIV:1309.0885;%%.

\bibitem{Mason:2013sva}
L.~Mason and D.~Skinner, ``{Ambitwistor strings and the scattering
  equations},'' {\em JHEP} {\bf 1407} (2014) 048,
\href{http://www.arXiv.org/abs/1311.2564}{{\tt 1311.2564}}.
%%CITATION = ARXIV:1311.2564;%%.

\bibitem{Adamo:2017nia}
T.~Adamo, E.~Casali, L.~Mason, and S.~Nekovar, ``{Scattering on plane waves and
  the double copy},'' {\em Class. Quant. Grav.} {\bf 35} (2018), no.~1, 015004,
\href{http://www.arXiv.org/abs/1706.08925}{{\tt 1706.08925}}.
%%CITATION = ARXIV:1706.08925;%%.

\bibitem{Bahjat-Abbas:2017htu}
N.~Bahjat-Abbas, A.~Luna, and C.~D. White, ``{The Kerr-Schild double copy in
  curved spacetime},'' {\em JHEP} {\bf 12} (2017) 004,
\href{http://www.arXiv.org/abs/1710.01953}{{\tt 1710.01953}}.
%%CITATION = ARXIV:1710.01953;%%.

\bibitem{Carrillo-Gonzalez:2017iyj}
M.~Carrillo-González, R.~Penco, and M.~Trodden, ``{The classical double copy
  in maximally symmetric spacetimes},'' {\em JHEP} {\bf 04} (2018) 028,
\href{http://www.arXiv.org/abs/1711.01296}{{\tt 1711.01296}}.
%%CITATION = ARXIV:1711.01296;%%.

\bibitem{Lee:2018gxc}
K.~Lee, ``{Kerr-Schild Double Field Theory and Classical Double Copy},''
\href{http://www.arXiv.org/abs/1807.08443}{{\tt 1807.08443}}.
%%CITATION = ARXIV:1807.08443;%%.

\bibitem{Monteiro:2018xev}
R.~Monteiro, I.~Nicholson, and D.~O'Connell, ``{Spinor-helicity and the
  algebraic classification of higher-dimensional spacetimes},''
\href{http://www.arXiv.org/abs/1809.03906}{{\tt 1809.03906}}.
%%CITATION = ARXIV:1809.03906;%%.

\bibitem{Monteiro:2013rya}
R.~Monteiro and D.~O'Connell, ``{The Kinematic Algebras from the Scattering
  Equations},'' {\em JHEP} {\bf 1403} (2014) 110,
\href{http://www.arXiv.org/abs/1311.1151}{{\tt 1311.1151}}.
%%CITATION = ARXIV:1311.1151;%%.

\bibitem{Tod}
K.~P. Tod, ``Self‐dual {Kerr–Schild} metrics and null {Maxwell} fields,''
  {\em Journal of Mathematical Physics} {\bf 23} (1982), no.~6, 1147--1148,
  \href{http://www.arXiv.org/abs/https://doi.org/10.1063/1.525482}{{\tt
  https://doi.org/10.1063/1.525482}}.

\bibitem{Eguchi:1978xp}
T.~Eguchi and A.~J. Hanson, ``{Asymptotically Flat Selfdual Solutions to
  Euclidean Gravity},'' {\em Phys. Lett.} {\bf 74B} (1978)
249--251.
%%CITATION = PHLTA,74B,249;%%.

\bibitem{Eguchi:1978gw}
T.~Eguchi and A.~J. Hanson, ``{Selfdual Solutions to Euclidean Gravity},'' {\em
  Annals Phys.} {\bf 120} (1979)
82.
%%CITATION = APNYA,120,82;%%.

\bibitem{Eguchi:1979yx}
T.~Eguchi and A.~J. Hanson, ``{GRAVITATIONAL INSTANTONS},'' {\em Gen. Rel.
  Grav.} {\bf 11} (1979)
315--320.
%%CITATION = GRGVA,11,315;%%.

\bibitem{Belavin:1975fg}
A.~A. Belavin, A.~M. Polyakov, A.~S. Schwartz, and {\relax Yu}.~S. Tyupkin,
  ``{Pseudoparticle Solutions of the Yang-Mills Equations},'' {\em Phys. Lett.}
  {\bf B59} (1975) 85--87.
[,350(1975)].
%%CITATION = PHLTA,B59,85;%%.

\bibitem{Ortin:2015hya}
T.~Ortin, {\em {Gravity and Strings}}.
\newblock Cambridge Monographs on Mathematical Physics. Cambridge University
  Press,
2015.
\newblock
%%CITATION = INSPIRE-1383727;%%.

\bibitem{Lee:2012rb}
S.~Lee, R.~Roychowdhury, and H.~S. Yang, ``{Test of Emergent Gravity},'' {\em
  Phys. Rev.} {\bf D88} (2013) 086007,
\href{http://www.arXiv.org/abs/1211.0207}{{\tt 1211.0207}}.
%%CITATION = ARXIV:1211.0207;%%.

\bibitem{Lee:2018czs}
J.~Lee and H.~S. Yang, ``{Quantized Kähler Geometry and Quantum Gravity},''
  {\em J. Korean Phys. Soc.} {\bf 72} (2018), no.~12, 1421--1441,
\href{http://www.arXiv.org/abs/1804.09171}{{\tt 1804.09171}}.
%%CITATION = ARXIV:1804.09171;%%.

\bibitem{Taub}
A.~H. Taub, ``Empty space-times admitting a three parameter group of motions,''
  {\em Annals of Mathematics} {\bf 53} (1951), no.~3, pp. 472--490.

\bibitem{NUT}
E.~Newman, L.~Tamburino, and T.~Unti, ``Empty‐space generalization of the
  schwarzschild metric,'' {\em Journal of Mathematical Physics} {\bf 4} (1963),
  no.~7, 915--923.

\bibitem{Parkes:1992rz}
A.~Parkes, ``{A Cubic action for selfdual Yang-Mills},'' {\em Phys.Lett.} {\bf
  B286} (1992) 265--270,
\href{http://www.arXiv.org/abs/hep-th/9203074}{{\tt hep-th/9203074}}.
%%CITATION = HEP-TH/9203074;%%.

\bibitem{tHooft:1988wxy}
G.~'t~Hooft, ``{A Physical Interpretation of Gravitational Instantons},'' {\em
  Nucl. Phys.} {\bf B315} (1989)
517--527.
%%CITATION = NUPHA,B315,517;%%.

\bibitem{Eguchi:1980jx}
T.~Eguchi, P.~B. Gilkey, and A.~J. Hanson, ``{Gravitation, Gauge Theories and
  Differential Geometry},'' {\em Phys. Rept.} {\bf 66} (1980)
213.
%%CITATION = PRPLC,66,213;%%.

\bibitem{Gibbons:1979xm}
G.~W. Gibbons and S.~W. Hawking, ``{Classification of Gravitational Instanton
  Symmetries},'' {\em Commun. Math. Phys.} {\bf 66} (1979)
291--310.
%%CITATION = CMPHA,66,291;%%.

\bibitem{Burnett-Stuart}
G.~Burnett-Stuart, ``{Sparling-Tod Metric = Eguchi-Hanson},'' {\em Twistor
  Newsletter} {\bf 9} (1979) 6--8.

\bibitem{Dirac:1931kp}
P.~A.~M. Dirac, ``{Quantized Singularities in the Electromagnetic Field},''
  {\em Proc. Roy. Soc. Lond.} {\bf A133} (1931) 60--72.
[,278(1931)].
%%CITATION = PRSLA,A133,60;%%.

\bibitem{White:2016jzc}
C.~D. White, ``{Exact solutions for the biadjoint scalar field},'' {\em Phys.
  Lett.} {\bf B763} (2016) 365--369,
\href{http://www.arXiv.org/abs/1606.04724}{{\tt 1606.04724}}.
%%CITATION = ARXIV:1606.04724;%%.

\bibitem{DeSmet:2017rve}
P.-J. De~Smet and C.~D. White, ``{Extended solutions for the biadjoint scalar
  field},'' {\em Phys. Lett.} {\bf B775} (2017) 163--167,
\href{http://www.arXiv.org/abs/1708.01103}{{\tt 1708.01103}}.
%%CITATION = ARXIV:1708.01103;%%.

\bibitem{tHooft:1976snw}
G.~'t~Hooft, ``{Computation of the Quantum Effects Due to a Four-Dimensional
  Pseudoparticle},'' {\em Phys. Rev.} {\bf D14} (1976) 3432--3450.
[,70(1976)].
%%CITATION = PHRVA,D14,3432;%%.

\end{thebibliography}\endgroup
\end{document}